\documentclass[aps,prb,twocolumn,floatfix,groupedaddress]{revtex4}
\usepackage{latexsym} \usepackage{amsmath} \usepackage{dcolumn}
\usepackage{bm} \usepackage{graphicx} \usepackage{epsfig}
\graphicspath{{./Figures/}}

\begin{document}


 \title{Real-space Green's function approach for intrinsic losses in x-ray spectra} 



\email[]{jjr@uw.edu}

\author{J. J. Kas} \affiliation{Dept.\ of Physics, Univ.\ of
Washington Seattle, WA 98195}
\author{J. J. Rehr} \affiliation{Dept.\ of Physics, Univ.\ of
Washington Seattle, WA 98195}


\date{\today}

\begin{abstract}

Intrinsic inelastic losses in x-ray spectra originate from excitations in an interacting electron  system due to a suddenly created core-hole. These losses characterize the  features observed in x-ray photoemission spectra (XPS), as well as many-body effects such as satellites and edge-singularities in x-ray absorption spectra (XAS). However, they are usually neglected in practical calculations.  As shown by Langreth these losses can be treated within linear response in terms of a cumulant Green's function in momentum space. 
Here we present a complementary {\it ab initio} real-space Green's function (RSGF) generalization of the 
Langreth cumulant in terms of the dynamically
screened core-hole interaction  $W_c(\omega)$ 
and the independent particle response function.  We find that the cumulant kernel $\beta(\omega)$
is analogous to XAS, but with the transition
operator replaced by the core-hole potential with monopole selection rules. The behavior
reflects the analytic structure of the loss function, with peaks near the zeros of the dielectric function, consistent with delocalized quasi-boson excitations.
The approach  simplifies when $W_c(\omega)$ is localized and spherically symmetric. 
Illustrative results and comparisons  are presented for the electron gas, sodium, and some early transition metal compounds.
\end{abstract}


\maketitle

\section{Introduction}
Intrinsic losses in x-ray spectra are  fundamental to the photoabsorption process.\cite{rehralb} They originate from the dynamic response of the system to a suddenly created core-hole, leading to dynamic screening by local fields and inelastic losses.
This transient  response is responsible for observable effects in   x-ray photoemission spectra (XPS) and x-ray absorption specta (XAS). These include satellites due to quasi-bosonic excitations such as   plasmons, charge-transfer, and shake-processes, as well as particle-hole excitations responsible for edge-singularity effects. These features are signatures of electronic correlation  beyond the independent particle approximation.\cite{zhou-pnas}   Various theoretical techniques have been developed for treating these losses, including plasmon models, quasi-boson approximations, fluctuation potentials, 
determinantal approaches, dynamical-mean-field theories, configuration-interaction methods, and
coupled-cluster approaches.\cite{langreth70,hedin-bard,hedin99rev,campbell,fujikawa,lgh,hmi,klevak14,NC,tzavala,prendergast,biermann2,bagus2022}  Recently cumulant Green's function methods have been developed\cite{Kas2015} based on a real-space real-time (RSRT) generalization  of the Langreth cumulant.\cite{langreth70} 
While the approach gives good results for the satellites observed in  XPS,  even for moderately correlated systems such as transition metal oxides,\cite{Kas2015,woicik1,woicik2} it   depends on computationally demanding real-time time-dependent density functional theory (TDDFT) calculations of the density-density response function. 
Thus despite these advances, quantitative calculations remain challenging, and intrinsic losses are usually neglected in current calculations of x-ray spectra.


In an effort to facilitate these calculations, we present here an   {\it ab initio} real-space Green's function (RSGF) generalization of the Langreth cumulant complementary to the RSRT approach, in which the  calculations are carried out using a discrete site-radial coordinate basis. The formalism of the cumulant kernel $\beta(\omega)$ is analogous to   
x-ray absorption spectra $\mu(\omega)$, except that the transition operator is replaced with the core-hole potential $V_c({\bf r})$, and the transitions are between valence and conduction states with monopole selection rules. 
  The generalized RSGF approach thereby permits 
calculations of many-body effects in x-ray spectra   in parallel with RSGF calculations of XAS.\cite{rehralb,feff10} Several  representations of $\beta(\omega)$ are  derived, which are useful  in the analysis and comparison with other approximations. For example, we show that $\beta(\omega)$ can be expressed either in terms of the bare core-hole potential $V_c({\bf r})$ and the full density response function $\chi({\bf r},{\bf r}',\omega)$, or the dynamically screened core hole potential $W_c({\bf r},\omega)$
and the independent particle response function $\chi^0({\bf r},{\bf r}',\omega)$. In adddition, we derive the link between the Langreth cumulant, and the commonly used approximation based on the $GW$ self-energy.\cite{aryasetiawan,guzzo, hedin99rev}
The potential $W_c({\bf r},\omega)$ is a key quantity of interest in this work. However,  its real-space behavior does not appear to have been extensively studied heretofore.   This quantity  
  is closely related to the dynamically screened Coulomb interaction $W({\bf r},{\bf r}',\omega)$ used e.g., 
in Hedin's $GW$ approximation for the one-electron self-energy.\cite{hedin99rev}  
The RSGF approach simplifies when  $W_c({\bf r},\omega)$ is  well localized and spherically symmetric. 
This leads to a local model for the cumulant kernel
on a 1-$d$ radial grid. 
  The local approach is tested with calculations for the homogeneous electron gas (HEG), and
  illustrative results are  presented   for nearly-free-electron systems and early $3$d transition metal compounds. 
   We   find that the local model provides a good approximation for $\beta(\omega)$. The model also accounts for the Anderson edge-singularity exponent in metals. 
 The behavior of the cumulant kernel 
  reflects the analytic structure of the loss function, with pronounced peaks near the zeros of the dielectric function. This structure is consistent with interpretations of intrinsic excitations in terms of  plasmons or charge-transfer excitations.
  
  
  The remainder of this paper is organized as follows.    Sec.\ II.\ summarizes the Langreth cumulant  and the RSGF and RSRT generalizations. Sec.\ III. and IV. respectively describe the calculation details and results for the HEG and charge-transfer systems.   Finally Sec.\ V. contains a summary and conclusions.  
  
  \section{Theory}
  
  \subsection{Cumulant Green's function and x-ray spectra}
  
    Intrinsic inelastic losses in x-ray spectra including particle-hole, plasmons, shake-up, etc., are characterized by features in the core-hole spectral function   
    $A(\omega)=\Sigma_n |S_n|^2 \delta(\omega-\varepsilon_n)$,
    where $S_n$ is the amplitude for an excitation of energy $\varepsilon_n$ due to the creation of the core-hole.  Equivalently $A_c(\omega)$ is given by  the Fourier transform of the core-hole Green's function, i.e., the one-particle Green's function with a deep  core-hole $c$ created at $t=0$  $g_c(t) =\langle 0|c_c^{\dagger} e^{iHt} c_c|0\rangle \theta(t)$,
\begin{equation}
    A_c(\omega) = -\frac{1}{\pi} {\rm Im} \int dt \, e^{-i\omega t} g_c(t).
    \label{spectralfn}
\end{equation}
Here $H$ is the Hamiltonian of the system while $c^{\dagger}_c$ and $c_c$ are creation and annihilation operators, respectively. 
The core-level XPS photocurrent $J_k(\omega) \sim A_c(\omega)$ is directly related to the spectral function,  which describes both the asymmetry of the quasi-particle peak and satellites in the spectra.  Intrinsic losses in
x-ray absorption spectra (XAS) $\mu(\omega)$ and related spectra (e.g., EELS) from   deep core-levels\cite{campbell}   can be expressed in terms of a convolution of $A_c(\omega)$ and the single (or quasiparticle) XAS $\mu_1(\omega)$\cite{campbell} 
  \begin{equation}
    \mu(\omega)=\int d\omega'\, A_c(\omega')\mu_1(\omega-\omega').
    \label{xasconv}
\end{equation}
This accounts for effects such as satellites and the reduction factor $S_0^2$ in the XAS fine structure.\cite{rehralb}
Here and below, unless otherwise noted, we use atomic units $e=\hbar=m=1$ with distances in Bohr = 0.529 \AA\ and energies in Hartrees = 27.2 eV.
A cumulant Green's function, which is a pure exponential in the time-domain, is particularly appropriate for the treatment of intrinsic losses,   
 \begin{equation}
      g_c(t) = g_c^0(t) e^{C_c(t)},
 \end{equation} 
where $g^0_c(t)=e^{i\epsilon_c t}$ is the independent particle Green's function for a given core-level $c$, and  $C_c(t)$  is the cumulant. This representation naturally separates single (or quasi-particle) and many-particle aspects of the final-state of the system with a deep core hole following photoabsorption, where many-body effects are embedded in the cumulant.   It is convenient in the interpretation to represent the cumulant in Landau form,\cite{landau44}
\begin{equation}
C_c(t) = \int d\omega\, \beta(\omega) \frac {e^{i\omega t} 
- i\omega t -1}{\omega^2}.
\label{eq:landau}
\end{equation}
 where the  {\it cumulant kernel} $\beta(\omega)$  characterizes the strength of the excitations at a given excitation energy $\omega$. 
This representation yields a normalized spectral function, with a quasi-particle renormalization constant  $Z=\exp(-a)$, where $a=\int d\omega\, \beta(\omega)/\omega^2$  is a dimensionless measure of correlation strength, and $\Delta=\int d\omega \beta(\omega)/\omega$
 is the relaxation energy shift of the core-level.\cite{hedin99rev} 
 \cite{hedin99rev} 
 
\subsection{Langreth cumulant} 

 For a deep core-hole coupled to the interacting electron gas  Langreth showed that within  linear response,      the  intrinsic inelastic losses can  be treated in terms of a cumulant Green's function,  with a  cumulant kernel in frequency and momentum space given by
\begin{equation}
\beta(\omega) = - \sum_q |V_q|^2 S({\bf q},\omega).
\label{langrethbeta}
\end{equation}
Here $V_q$ is the Fourier transform of the core-hole potential
and $S({\bf q},\omega)$ is the dynamic structure factor,  which is directly related to 
time-Fourier transform of the density-density correlation function $\chi({\bf q},\omega)$ and the loss function $L({\bf q},\omega) =   -{\rm Im}\, \epsilon^{-1}({\bf q},\omega)$, i.e., 
\begin{eqnarray}
 S({\bf q},\omega) &\equiv& - \frac{1}{\pi} {\rm Im}\, \chi({\bf q},\omega)\theta(\omega)  
 \nonumber \\
 &=&  -\frac{1}{\pi v_q} {\rm Im}\, \epsilon^{-1}({\bf q},\omega)\theta(\omega),   \\
\chi({\bf q},\omega) &=&   \int dt\, e^{i\omega t} \langle \rho_{\bf q}(t) \rho_{\bf q} (0)\rangle.
\end{eqnarray}
The response function $\chi({\bf q},\omega)$ can be expressed in terms of the non-interacting response $\chi^0({\bf q},\omega)$ using the relation $\chi=\chi^0+ \chi^0 K \chi$. Here the particle-hole interaction kernel $K$ within TDDFT   is given by $K=v+f_{xc}$, where  $v=4\pi/q^2$ is the bare Coulomb interaction, and $f_{xc} = \delta v_{xc}[n]/\delta n$ is the TDDFT kernel; this   is obtained from the  exchange-correlation potential  used in the definition of the independent particle response function $\chi^0(\omega)$.

Although Langreth's expression for $\beta(\omega)$ in Eq.\ (\ref{langrethbeta}) is elegant for its simplicity, calculations of the full density response function $\chi({\bf q},\omega)$   are challenging, being comparable to that for a particle-hole Green's function or the  Bethe-Salpeter equation (BSE). Moreover, the core-hole potential  $V_c({\bf r})$ has a long range Coulomb tail to deal with. On the other hand, the Thomas-Fermi screening length $r_{0} = 0.64\, r_s^{1/2}$ is short-ranged, so one may wonder to what extent a local approximation for the dynamically screened interaction might be applicable?  

To this end, we note that the cumulant kernel can be expressed equivalently in terms of the dynamically screened core-hole interactions $W_{\bf q}(\omega) = V_q/\epsilon({\bf q},\omega)$ and the independent particle response function $\chi^0({\bf q},\omega)$,  
\begin{eqnarray}
\beta(\omega) &=& \frac{1}{\pi}\sum_q |V_q|^2 {\rm Im}\, \chi({\bf q},\omega)\theta(\omega), \label{langrethV} \\
    &\equiv& \frac{1}{\pi}\sum_q |W_q(\omega)|^2 \left[{\rm Im}\, \chi^0({\bf q},\omega) +\right. \nonumber \\
    &~&\ + \left.|\chi^0({\bf q},\omega)|^2 {\rm Im} f_{xc} \right] \theta(\omega).
    \label{langrethKW}
\end{eqnarray}
This equivalence is  implicit in Langreth's analysis of the low energy behavior of $\beta(\omega)$ for the homogeneous electron gas (HEG) in the random phase approximation (RPA), where an adiabatic approximation is also valid $\epsilon({\bf q},\omega) \approx \epsilon({\bf q},0)$.
If the exchange correlation kernel $f_{xc}$ is taken to be real, as in typical implementations of TDDFT  or ignored as in   the RPA ($f_{xc}=0$), Eq.\ (\ref{langrethW}) reduces to 
\begin{equation}
    \beta(\omega) = \frac{1}{\pi}\sum_q |W_q(\omega)|^2 {\rm Im}\, \chi^0({\bf q},\omega)\theta(\omega).
    \label{langrethW}
\end{equation}
This result  in terms of the screened-core-hole potential 
$W_q(\omega)$ can be advantageous for practical calculations in inhomogeneous systems. For example, in the adiabatic approximation $W_q(\omega)\approx W_q(0)$, only a single matrix inversion is needed to obtain $\epsilon^{-1}(\omega=0)$, rather than an inversion at each frequency.

As a practical alternative to momentum-space,  calculations of the Langreth cumulant  
for inhomogeneous systems
have recently been carried out  by transforming to   real-space  and real-time (RSRT).\cite{Kas2015}   
 The approach first calculates the time-evolved density response $\delta \rho({\bf r},t)$ with the Yabana-Bertsch reformulation of TDDFT that builds in a DFT exchange-correlation
 kernel\cite{Yabana-Bertsch}        
\begin{equation} 
\delta\rho({\bf r} ,t)\equiv \int d^3 r'dt'\, \chi({\bf r},t;{\bf r}',t') V_c({\bf r'},t') .
\end{equation}
A Fourier transform then yields the cumulant kernel $\beta(\omega)$  
\begin{equation}
 {\beta(\omega)}  =  {\rm Re}\, \frac{\omega}{\pi} 
\int_0^{\infty} dt\, e^{-i\omega t} 
\int d^3 r\,   
V_c({\bf r}) \delta\rho({\bf r},t)\, \theta(\omega). 
\label{deltact}
\end{equation}
This approach has been shown to give good results for a number of systems.\cite{Kas2015}  However, the   method requires a computationally demanding long-time evolution of the density response using a large supercell, with Kohn-Sham DFT calculations at each time-step. In addition,  a post-processing convolution is needed for XAS calculations.  
  

\subsection{Real-space Green's function Cumulant}

Our primary goal in this work is to develop an alternative, real-space Green's function formulation of the Langreth cumulant and it's key ingredients. In particular we aim to investigate the behavior of the cumulant kernel $\beta(\omega)$ and the dynamically screened core-hole potential $W_c({\bf r}.\omega)$.
The  method is complementary to the  RSRT formulation but is based on a similar approach for the response function, and can be carried out in parallel with RSGF calculations of XAS.\cite{rehralb,kas2021feff10} 
Within the RPA, the RSGF formulation of $\beta(\omega)$ can be derived from the perhaps better known GW approximation to the cumulant, based on the core-level self-energy 
 $\Sigma_c(\omega)$,\cite{aryasetiawan,almbladh,guzzo} 
 \begin{equation}
     \beta(\omega) = \frac{1}{\pi}{\rm Im}\, \Sigma_{c}(\epsilon_c - \omega),
      \label{betaGW}
     \end{equation}
       where   
       $\Sigma_c(\omega) = \langle c| \Sigma(\omega) |c\rangle$.  
     This matrix element can be evaluated in real-space  using the GW approximation
     \begin{align}
           {\rm Im}\, \Sigma({\bf r},{\bf r'},\omega) =& -\sum_i^{occ}\psi_i({\bf r})\psi_i({\bf r'}) \times \nonumber \\
           &{\rm Im}[W({\bf r},{\bf r'},\epsilon_i-\omega)]\theta(\epsilon_i-\omega), 
     \end{align}
where $W({\bf r},{\bf r'},\omega) $
 is the dynamically screened Coulomb interaction. Within 
   the decoupling approximation,\cite{guzzo} the core-level wave-function is assumed to have no overlap with any other electrons, and Eq.\ (\ref{betaGW}) becomes
 \begin{eqnarray}
     \beta(\omega) &=& \frac{1}{\pi}{\rm Im}\!\!\int\!\! d^3r d^3 r' \rho_c({\bf r}) W({\bf r},{\bf r'},\omega) \rho_c({\bf r}') \theta(\omega)   \\
     &=&\frac{1}{\pi}{\rm Im}\!\!\int d^3r   \, \rho_c({\bf r})W_c({\bf r},\omega)\theta(\omega),
     \label{betase}
 \end{eqnarray}
where 
   $W_c({\bf r},\omega)= \int d^3 r'\, W({\bf r},{\bf r'},\omega) \rho_c({\bf r}') $ 
and $\rho_c({\bf r})=|\psi_c({\bf r})|^2$.
Then noting  that ${\rm Im}\,W_c= {\rm Im}\,[K \chi V_c]$ (indices suppressed for simplicity) and within the RPA ($K=v$), we obtain a  real-space  generalization  of 
the Langreth cumulant in Eq.\ (\ref{langrethV})  
\begin{equation}
     \beta(\omega) = \frac{1}{\pi}\int d^3r d^3r'\ V_c({\bf r}) V_c({\bf r'})\,{\rm Im}\,\chi({\bf r},{\bf r'},\omega)\theta(\omega).
     \label{betaVV}
 \end{equation}
This result is also equivalent to the RSRT expression in Eq.\ (\ref{deltact}). Alternatively,
 in analogy with  Eq.\ (\ref{langrethW}), and again within the RPA,
 $\beta(\omega)$   can    be expressed    in terms  of       $W_c({\bf r},\omega)$ and the independent particle 
 response function  
\begin{align}
\beta(\omega) = &\frac{1}{\pi} \int d^3 r d^3 r'\, W_c^*({\bf r},\omega)W_c({\bf r}',\omega) \times \nonumber \\
& \times {\rm Im}\, \chi^0({\bf r},{\bf r}',\omega)\theta(\omega).
\label{betaWW}
\end{align}
While Eqs.\ (\ref{betase}-\ref{betaWW}) are formally equivalent within the RPA, they differ if the interaction kernel $K$ is complex (which is the case for most non-adiabatic kernels), in which case it is not obvious which approximation is best. Here we focus on the RPA expressions only, although a generalization to adiabatic $f_{xc}$ would be relatively simple. 

The static limit of Eq.\ (\ref{betaWW})  is interesting in itself. At frequencies well below  $\omega_p$, the core-hole potential is strongly screened beyond the screening length $r_0$ and nearly static. 
On expanding  $\chi({\bf r},{\bf r}',\omega)$ about $\omega=0$ and keeping only the leading terms, one obtains the adiabatic approximation  
\begin{equation}
\beta(\omega) \approx \frac{1}{\pi} \int d^3 r d^3 r'\, W_c^*({\bf r},0)W_c({\bf r}',0)\,
{\rm Im}\, \chi^0({\bf r},{\bf r}',\omega).  
\label{betaWW0}
\end{equation}
This limiting behavior is  similar to the adiabatic TDDFT approximation for 
XAS,\cite{tddft-bse} but with the replacement of the dipole operator $d({\bf r})$ by the statically screened core-hole potential $W_c({\bf r},0)$. This potential is also used in calculations of XAS to approximate the particle-hole interaction, and is similar to the {\it final state rule} approximation for the static core-hole potential. 


The transformation in Eq.\ (\ref{betase}) emphasizes the localization of $\beta(\omega)$, which only depends on the  imaginary part of the screened core-hole potential ${\rm Im}\, W_c({\bf r},\omega)$ over the range of the core density $\rho_c({\bf r})$. %
 The above identities also illustrate the connection between the local and extended behavior of the dynamical response function.  This connection is analogous, e.g., to the origin of fine-structure in XAS, where back-scattering is responsible  for the fine-structure in photoelectron wave-function at the origin.  

Not surprisingly, since both quantities are physically related to dielectric response,  calculations of $\beta(\omega)$ are formally  similar to those for XAS.\cite{prange_opcons}
\begin{equation}
\mu(\omega)= \frac{4\pi}{V}  \int d^3 r d^3 r'\, d({\bf r})d({\bf r}')\,
{\rm Im}\, \chi({\bf r},{\bf r}',\omega), 
\label{eq:xas}
   \end{equation}
where ${d}(\bf r)$ is the (e.g., dipole) transition operator. Analogous expressions 
have been derived for TDDFT  approximations of  atomic  polarizabilities \cite{zangsov,zaremba} and optical absorption spectra.\cite{tddft-bse}  The quasi-particle XAS $\mu_1(\omega)$ in  our RSGF calculations\cite{rehralb,kas2021feff10} is obtained by evaluating $\mu(\omega)$ using the final-state rule, i.e., with a screened core-hole in the final state, which corresponds to the adiabatic approximation of Eq.\ (\ref{betaWW0}). Physically the absorption in $\mu(\omega)$ can be viewed in terms of the damped oscillating electric dipole moment of the system induced by an external electric dipole potential oscillating at frequency $\omega$. The kernel $\beta(\omega)$ can be viewed similarly, except that the dipole potential $d({\bf r})$ is replaced by the core-hole potential $V_c({\bf r})$. Thus a major difference is that the suddenly turned on core-hole induces a damped oscillating monopole response field about the absorbing atom.



 Formally $W_c({\bf r},\omega)$  
can be  calculated in terms of the inverse TDDFT dielectric matrix in real-space\cite{zangsov}  
\begin{eqnarray}
W_c({\bf r},\omega) &=& \int d^3 r'\, \epsilon ^{-1}({\bf r},{\bf r}',\omega) V_c({\bf r} '),
\label{screenedW} \\
   \epsilon({\bf r},{\bf r}',\omega) &=& \delta({\bf r},{\bf r}')\!-\!\!\int\! d^3r'' K({\bf r},{\bf r}'')\chi^0({\bf r}'',{\bf r}',\omega) ,
   \label{dielectricfn}
\end{eqnarray}
Related expressions for $W_c({\bf r},\omega)$ 
in atoms  have also been reported.\cite{shirleyatom}
Alternatively  $W_c({\bf r},\omega)$ can be obtained by iterating the integral equation $W_c= V_c  + K \chi^0 W_c$
to self-consistency.\cite{zangsov} 
Yet another tack  is the use of {\it fluctuation potentials}, e.g., in the quasi-boson approach.\cite{hedin99rev}  These  
  are obtained by diagonalizing the dielectric matrix 
 using an eigenvalue problem for each frequency $\omega$,   
\begin{equation}
 \int d^3 r' \epsilon({\bf r},{\bf r}',\omega) w_s({\bf r}',\omega) = \lambda_s(\omega) w_s({\bf r},\omega),  
  \label{eigenprob}
\end{equation}
The eigenfunctions $w_s(\omega)$ are the same as those for $K\chi^0$, which has eigenvalues $\kappa_s(\omega)=1-\lambda_s(\omega)$. This approach is particularly useful near the quasi-bosonic resonances $\omega_s=\omega_p$ where ${\rm Re}\, \lambda_p(\omega)$ crosses zero and matrix inversion can be numerically unstable.  The corresponding excitation energies $\omega_p$ and  eigenfunctions $w_p({\bf r},\omega)$  define the   {\it fluctuation potentials}.\cite{hedin99rev,hmi}
Close to $\omega_p$,   $\lambda_p(\omega)$ varies linearly,
\begin{equation}
\lambda_{p}(\omega) \approx \lambda'_p (\omega-\omega_p)  + i\Gamma_p,    
\end{equation}
where $\lambda'_p = d[ {\rm Re}\, \lambda_p(\omega)]/{d\omega}|_{\omega_p}$.
This approximation  yields a Lorentzian behavior for the   quasi-bosonic   peaks in  $\beta(\omega)$ of  width $\gamma_p = \Gamma_p/\lambda_p'$, which also limits the range of the density fluctuations and the screened potential  $w_p({\bf r},\omega_p$).
Nevertheless,  we have found that an explicit matrix inversion on a finite real-space basis (see below) usually converges well for all frequencies due to the finite
imaginary part of the dielectric screening $\epsilon(\omega)$.
Formally, the Lindhard expression for the independent particle response function on the real-axis can be expressed in terms of the one-particle Green's function \cite{prange_opcons}
\begin{align}
\chi^0({\bf r}, {\bf r}', \omega)  =   
  &2\int^{E_F}_{-\infty} \hspace{-1em} dE\,
  [\rho({\bf r}, {\bf r}',E)  G({\bf r}', {\bf r}, E+\omega)
   \nonumber \\
   &+ \rho({\bf r'}, {\bf r},E) G^*({\bf r} , {\bf r}', E-\omega) ], \\
   \frac{-1}{\pi}{\rm Im}\,\chi^0({\bf r}, {\bf r}', \omega)  &= 2\int_{E_F-\omega}^{E_F} dE\,
   \rho({\bf r}, {\bf r}',E)\rho({\bf r}', {\bf r}, E+\omega).
\end{align}
Here $\rho({\bf r'}, {\bf r},E)$ is the spectral density of the
one-particle Green's function  which under the integrals can be replaced with $(-1/\pi){\rm Im}\, G({\bf r'}, {\bf r},E)$ (cf. Eq.\ (33) and (34) of  Ref\ \onlinecite{zangsov}). The convolution  in  ${\rm Im}\, \chi^0(\omega)$ defines a particle-hole spectral function. The calculation of $W_c({\bf r},\omega)$ thus requires both Re and ${\rm Im}\, \chi^0({\bf r},{\bf r}',\omega)$, and
hence  expressions for both $\rho({\bf r}, {\bf r}',E)$ and $G({\bf r},{\bf r}',\omega)$. Explicit derivations and algorithms for   these functions 
are given in Ref.\ \onlinecite{prange_opcons}, along with algorithms for calculating   XAS and optical response, based on summations over a finite cluster surrounding the absorbing atom. Below, we show how $\chi^0({\bf r},{\bf r}',\omega)$ and $\chi({\bf r},{\bf r}',\omega)$ can be calculated within a generalized real-space multiple-scattering Green's function (RSGF) approach using a discrete site-radial coordinate basis.


\subsection{Real-Space Multiple Scattering Formalism}

The key ingredients in the RSGF calculation of  $W_c({\bf r},\omega)$  are the bare response function $\chi^0({\bf r},{\bf r}',\omega)$ 
which is defined in Eq.\ (23) in terms of the independent particle Green's functions $G({\bf r},{\bf r}',E)$,  and the interaction kernel $K = v+f_{xc}$ which here is approximated by the RPA $K= v$. 
The most demanding step  is the inversion of the non-local dielectric matrix $\epsilon({\bf r},{\bf r}',\omega)$ on a 3-$d$ grid, which can be computationally formidable.   
In order to simplify the calculation, we employ a generalization of  real-space multiple-scattering (RSMS) formalism  with a discrete site-radial coordinate basis,  analogous to that developed for x-ray spectra and optical
response.\cite{rehralb,prange_opcons}
In this approach, space is partitioned  into cells $i$ with cell boundaries defined by $\Theta_i({\bf r})=1\ (0)$ for points inside (outside) a given cell. Formally the cell boundaries should be defined by Voronoi or equivalent  partitioning. However, for simplicity, the cells are taken to be Norman spheres (i.e., spheres of charge neutrality).\cite{norman,prange_opcons}  
Thus $\Theta_i({\bf r})=\theta(r_N-r_i)$, where $\theta(r)$ is the unit step function, and $r_N$ denotes the Norman radius. The points within each cell
are then represented in spherical polar coordinates ${\bf r}_i = (r_i,\Omega_i)$. In addition the radial coordinates are typically defined on a discretized logarithmic grid, e.g.,  $r_{i,n}=\exp(-x_0 + n \delta x)$, as in SCF atomic calculations,\cite{Descl} while the angular dependence  is represented by spherical harmonics $Y_{L_i}(\hat r_i)$, where $L_i = (l_i,m_i)$. Thus the spatial points ${\bf r}$ are represented by the discrete indices  $I = (i, n_i,L_i)$, which have a finite volume $\Delta_i = 4\pi r_i^2 \delta_i$, with $\delta_i\approx r_i\delta x$ being the radial grid spacing (e.g., $x_0 =-8.8$ and $\delta x$ = 0.05).  This construction conserves charge and simplifies volume integration, i.e., for a given volume
$\int d^3 r = \sum_i \int^{r_N} r_i^2 d r_i d\Omega_i$.
With this representation, functions of coordinates  
$F({\bf r})\rightarrow F_I$ are represented by vectors, and $F({\bf r},{\bf r}')\rightarrow F_{I,J}$ by matrices in $I$ and $J$ of rank $N_r N_i(l_{max}+1)^2$, where $l_{max}$ denotes the maximum angular momentum used in the calculation.

This contruction is then implemented within the standard RSMS theory, which assumes spherical symmetry of the scattering potentials within each cell.  The independent particle Green's functions are given by,\cite{rehralb}
\begin{align}
    G({\bf r}, {\bf r'}, E) =& -2k\sum_{iLjL'}\left[\delta_{ij}\delta_{LL'}R_{iL}({\bf r}_{i<}{\bf r}_{i>}) +\right. \nonumber \\
    &\left. R_{iL}({\bf r_i}) \tilde G_{iLjL'}(E)R_{jL'}({\bf r}'_j)\right]\Theta_i({\bf r})\Theta_j({\bf r'}) \nonumber \\
    =&\sum_{iLjL'} \left[G^{c}_{iL}({\bf r}_i,{\bf r}'_j,E)\delta_{ij}\delta_{LL'}\right. +\nonumber \\  &\left. G^{sc}_{iLjL'}({\bf r}_i,{\bf r}'_j,E)\right]\Theta_i({\bf r})\Theta_j({\bf r}').
    \label{rsmsgf_full}
\end{align}
Here 
$R_{iL}({\bf r})=i^l(R_{il}(r)/r)Y_L({\bf \hat r})$ and $H_{iL}({\bf r})=i^l(H_l(r)/r)Y_L({\bf \hat r})$ are the regular and irregular solutions to the single site Dirac equation for the atom at site $i$, ${\bf r_i=r-R_{i}}$ is the position relative to the center of the $i^{th}$ cell, and $\tilde G_{iLjL'}(E)$ is the multiple scattering matrix.\cite{prange_opcons}
The radial wave-functions $R_L({\bf r},E)= i^l (R_l(r)/r) Y_L(\hat r)$ are scattering-state normalized such that $R_l(r)\rightarrow \sin(kr -l\pi/2+\delta_l)$  beyond the muffin-tin radius $r_{mt}$,\cite{alex} 
where $\delta_l$ are partial wave phase shifts and $k=\sqrt{2 E}$ is the photoelectron momentum.  Thus, 
\begin{equation}
    G({\bf r,\bf r'},E) = \sum_{iLjL'}Y_L^*(\hat r_i)G_{iLjL'}({r}_i,{ r}'_j)Y_{L'}(\hat {r'_j}),
\end{equation}
where the cell functions $\Theta_i$ and $\Theta_j$ have been absorbed into the definition of $G_{iLjL'}(r_i,r'_j,E)$. 
On the real axis, the density matrix ${\rho = {\rm Im}\, G}$ can be expressed similarly, 
\begin{align}
    \rho({\bf r},{\bf r'}) &= \sum_{iLjL'}R_{iL}({\bf r_i})\tilde\rho_{iLjL'}(E)R_{jL'}({\bf r'_j})\Theta_i({\bf r})\Theta_j({\bf r'}), \nonumber \\
    &=\sum_{iLjL'}Y_{L}(\hat r)\rho_{iLjL'}(r_i,r'_j,E)Y_{L'}(\hat {r'}).
\end{align}
Additional details of the method and algorithms for calculations of the quantities involved in Eq.\ (\ref{rsmsgf_full}) can be found elsewhere.\cite{rehralb,prange_opcons} 
With the above representation of the Green's function and density matrix, the bare response function $\chi^0(\omega)$ can   be expressed similarly in a site, radial-coordinate, and angular momentum basis,
\begin{align}
    \chi^0({\bf r},{\bf r}',\omega) = \sum_{iLjL'}Y_L(\hat r_i)\chi^0_{iLjL'}({ r}_i,{  r}'_j,\omega) Y_{L'}({\hat r}'_{j}).
\end{align}
 Likewise,  we expand the interaction kernel $K$ in spherical harmonics about each cell center, and for simplicity given the near spherical symmetry of the core-hole,  keep only the spherical terms $L=L'=0$ . If we ignore the TDDFT contribution   
$f_{xc}=\partial V_{xc}(\rho)/\partial \rho $,  
as in the RPA, the kernel becomes
\begin{align}
    K({\bf r},{\bf r}') &\approx \sum_{ij}Y_{00}(\hat r_i)K_{i0j0}(r_i,r'_j)Y_{00}(\hat r'_j) \Theta_i({\bf r})\Theta_j({\bf r'}), \nonumber \\ &K_{i0j0}(r_i,r'_j) = \left[\frac{4\pi}{r^{i}_>}\delta_{ij} + \frac{4\pi}{R_{ij}}(1-\delta_{ij})\right],
\end{align}
where $R_{ij} = 1/|{\bf R}_i-{\bf R}_j|$, and $r_{i>}=\max(r_i,r'_i)$. 

As a further simplification, we assume here that the deep core-hole density and potential, and hence the screened core-hole potential are spherically symmetric about the central site,
\begin{align}
     \rho_c({\bf r}) &= \rho_{c0}(r)Y_{00}({\bf \hat r}), \nonumber\\
     V_c({\bf r}) &= V_{c0}(r)Y_{00}({\bf \hat r}), \nonumber \\
     W_c({\bf r},\omega) &= w_{c}(r,\omega)Y_{00}({\bf \hat r}).
 \end{align}
Then  $W_c({\bf r},\omega)$ as well as $\beta(\omega)$ depend only on the spherical $L=L'=0$ components of the bare response function.
  The matrix inversion of  ${\bf {\boldsymbol{\epsilon = 1 - K \chi_0}}}$  can then be expressed in terms of a matrix inverse in the site and radial coordinates 
\begin{equation}
     \boldsymbol{\chi}=\boldsymbol{\chi^0}[{\bf 1-K}\boldsymbol{\chi^0}]^{-1}.
     \label{matrix_chi}
\end{equation}
From Eq.\ (24),  the matrix elements  $\boldsymbol{\chi^0}_{IJ}$ in the discrete  basis $I=(i,r_i)$ and $J=(j,r'_j)$ are 
\begin{align}
 &\boldsymbol{\chi^0}_{I,J}(\omega) = \frac{1}{2\pi}\int_{-\infty}^{E_F}\sum_{LL'} \Delta_i^{1/2} \Delta_j^{1/2} \rho_{iLjL'}(r_i,r'_j,E) \times \nonumber \\
&\ \left[G_{jL'iL}(r'_j,r_i,E+\omega)
+ G^*_{jL'iL}(r'_j,r_i,E-\omega)\right]. 
\end{align}
Examining the above equations and Eq.\ (\ref{rsmsgf_full}), we   see that site 
off-diagonal terms $i\neq j$ contribute to the full response function in several ways: i) off-diagonal terms in $\chi^0$ corresponding to the free propagation of a particle-hole from one site to another; ii) off-diagonal terms in the interaction kernel $K$, corresponding to the coupling of  a particle-hole state at site $i$ with another at site $j$; iii) a combination of these two, corresponding to the propagation of a particle-hole state from one site to an intermediate site, then scattering into another particle-hole state at a third site; and iv) the Green's function itself has contributions due to scattering of the single particle (photoelectron or valence hole) from neighboring sites. 

To summarize, the expressions for $\beta(\omega)$ in the discrete site-radial coordinate basis are  
\begin{align}
\beta(\omega)    &= {\rm Im}\left[\boldsymbol{\rho_{c0}}{\bf W_{c0}}\right]\theta(\omega), \nonumber \\
&= {\rm Im}[{\bf V_{c0}\chi V_{c0}}]\theta(\omega),\nonumber \\
    &={\rm Im}\left[{\bf W_{c0}^{*} }\boldsymbol{\chi^0}{\bf W_{c0}} \right]\theta(\omega),
    \label{beta_matrixform}
\end{align}
where the vectors are given by
\begin{align}
    \boldsymbol{\rho_{c0} } &= \boldsymbol{\rho_{cI}}\delta_{i0}, \nonumber \\
    {\bf V_{c0}} &= {\bf K} \boldsymbol{\rho_{c0}} , \nonumber \\
    {\bf W_{c0}} &= ({\bf 1-K}\boldsymbol{\chi^0})^{-1}{\bf V_{c0}  }.
    \label{matrix_beta}
\end{align}
The last line in Eq.\ (\ref{beta_matrixform}) can also be recast in terms of matrix elements of the screened core-hole potential 
\begin{equation}
    W_{iL}(E,\omega)=\int^{r_{N}} dr R_{il}(r_i,E) w_c(r,\omega) R_{il}(r_i,E+\omega).
\end{equation}
 This reveals a striking resemblance to the RSMS theory of XAS,\cite{rehralb} or more precisely, that of optical response,\cite{prange_opcons}
 the difference being that the dipole transition operator ${  d(r)}$ is replaced by the dynamically screened core-hole potential $w_c(r,\omega)$,
\begin{align}
    \beta(\omega) =  \int_{E_F}^{E_F+\omega} &dE\sum_{iLjL'} W_{iL}^*(E,\omega) \rho_{iLjL'}(E) \times \nonumber \\
    &\rho_{iLjL'}(E+\omega)W_{jL'}(E,\omega).
\end{align}

\subsection{Local RSGF cumulant approximation}

The first line in 
Eq.\ (\ref{beta_matrixform}) shows that the screened core-hole potential is only needed locally about the absorbing atom, since it is multiplied by the density of the core-orbital. For example, for the $1s$ state of Na, the required range is only $\sim 1/11$ Bohr. This suggests that it may be possible, at least in some systems, to treat the problem locally, neglecting site 
off-diagonal terms in $K$ and $\chi^0$, while in other systems it may be sufficient to approximate the contribution from off diagonal terms. To do this, the Green's functions in $\chi^0$ are taken to be those of an atom embedded in an electron gas at the interstitial density $r_s$, for points $r, r'$ outside the Norman radius of the absorbing atom, while the full Green's function is used within the Norman radius. Thus outside the absorbing cell, the Green's function is approximated simply in terms of phase-shifted spherical Bessel functions centered about the absorbing site. The dielectric matrix $\boldsymbol{\epsilon}=\bf {1 -K} \boldsymbol{\chi^0}$ is then inverted on a single radial grid, which greatly simplifies the calculations. In this case the matrix equations Eq.\ (\ref{matrix_chi}) and (\ref{matrix_beta}) still apply, but are limited to $i=0$, and the radius $R_{\rm max}$ defining the central cell is treated as a convergence parameter. The equation for $\beta(\omega)$ then becomes  
\begin{equation}
\beta(\omega) = \frac{1}{\pi}\int_0^{R_{\rm max}}dr\, \rho_c^0(r)w_c^0(r,\omega).
\end{equation}
If  we further assume that the Green's function has the structure of that of a spherical system, i.e., $G_{L,L'}=G_{l}\, \delta_{LL'}$, one can express $\beta(\omega)$ in terms of a joint density of states,
[{\it cf.} Eq.\ (32) of Prange et al.],\cite{prange_opcons}  
\begin{align}
 \beta(\omega) =  2\sum_{l} 
 (2l+1) &\int_{E_F-\omega}^{E_F}\hspace{-1em} dE\, |W_{l}(E,\omega)|^2
\rho_{l}(E) \times \nonumber \\
&\rho_{l}(E+\omega)\theta(\omega),
\label{betaspherical}
\end{align}
where  $\rho_l(E)$ is  the angular-momentum projected density of states (LDOS) for a given $l$ calculated at the central atom, and the factor of 2 accounts for spin degeneracy.

With spherical symmetry of the core-hole density, calculations   of $\beta(\omega)$   reduce 
to  a form with only radial coordinates.
The spherical part of the independent particle response function
$\chi^0({ r}, {  r}', \omega)|_{l=0}$, \cite{prange_opcons} is
\begin{eqnarray}
  \chi^0_{l=0}({  r}, {  r}', \omega)   &=&  
 \frac{1}{2\pi} \int_{-\infty}^{E_F} \hspace{-1em} dE\, \sum_l (2l+1) 
  \rho_l({ r}, { r}',E) \times \nonumber \\
  \times\, && \hspace{-2em}[G_l({ r}, { r}', E+\omega) +
  G_l^*({ r} , { r}', E-\omega) ]. 
\end{eqnarray}
Here we have used the symmetry and reality of
$\rho_l(r,r',E)$ and $G_l(r,r',E)$.
In order to stabilize this expression, we split the two terms of the energy integral as follows
\begin{eqnarray}
\chi^0_{l=0}(r,r',\omega) &=& \frac{1}{2\pi}\sum_l (2l+1)  [\chi^0_l(I) + \chi^0_l(II)], \\
   \chi^0_l(I)&=& \int_{E_F-\omega}^{E_F}\hspace{-1em} dE\ \rho_l(r,r',E)G_l(r,r',E+\omega), \nonumber \\
    \chi^0_l(II) &=& - \frac{1}{\pi}{\rm Im}\, \int^{E_F}\hspace{-1em}dE\, G_l(r,r',E)G_l(r,r',E-\omega).\nonumber 
\end{eqnarray}
The second  of the  above integrals can   be performed in the complex energy plane where the Green's function is smoother. 
Note  also  that these expressions implicitly include fine-structure in $G_l(r,r',\omega)$ from multiple-scattering from atoms   beyond the 
central atom. 
Finally the screened core hole potential $w_c(r,\omega)$ is defined  by  a the radial integral,
  \begin{equation}
 w_c(r,\omega)= \int dr' \, [1-K\chi^0(\omega)]^{-1}_{l=0}(r,r')   v_c(r'),
       \label{eq:wcr}
 \end{equation}
 which can be  calculated directly by matrix inversion in the radial coordinate basis for the single site $i=0$.
We  use this local RSGF approximation in all of our results shown below.

  


Eq.\ (\ref{betaspherical}) 
is also consistent with the Anderson edge-singularity exponent $\alpha$ for XAS.
For metallic systems, $\beta(\omega)\approx \alpha\ \omega$ is roughly linear in frequency for small $\omega$, reflecting the behavior of the joint density of states near the Fermi energy. The coefficient  $\alpha$ can  be determined from the zero frequency limit,   with    matrix elements given by their values at the Fermi level.        Since the screened core-hole potential is well localized and spherically symmetric near $\omega=0$,   
\begin{equation}
    \alpha  =  2\sum_l (2l+1) |W_l(E_F,\omega=0)|^2 \rho_l(E_F)^2.
   \end{equation}
    This   result is consistent with that derived by  Anderson, Noz\`ieres and De Dominicis\cite{ND}  
 $ \alpha = 2\Sigma_l (2l+1)  {(\delta_l/\pi)^2}$.   
This can be verified by noting that $\rho_l(E)  = d N_l/dE$, and $W_l(E_F)=\Delta E_l$
is the change in energy of levels $l$  due to the screened core-hole potential $W(r,0)$, so   $W_l(E_F)\rho_l(E_F) = \Delta N_l = \delta_l/\pi$, i.e., the screening charge in   level  $l$, as in the Friedel sum rule. This singular behavior shows up as an asymmetry in the main peak  of the XPS. The XAS has an additional contribution from the Mahan edge singularity exponent $\alpha_l=-2\delta_l/\pi$, where the photoelectron has local angular momentum $l$.\cite{rehr20} 
Even in insulators, the  low energy background terms tend to grow linearly beyond the gap, leading to an asymmetry in the quasi-particle peak, though without a true singularity.

\section{Details of the Calculations}

The RSGF formalism used here has been implemented within the FEFF10 code,\cite{feff10} with the RPA approximation for the TDDFT kernel  $K=1/|{\bf r}-{\bf r}'|$. The one-particle Green's functions are calculated using self-consistent-field (SCF) potentials, full-multiple-scattering,\cite{rehralb,kas2021feff10} and unless noted otherwise, the final-state-rule approximation with a screened core hole in the final state.   The RSGF calculations of $w_c(r,\omega)$ depend on the  radius $R_{max}$ beyond which the response functions are truncated, the maximum angular momentum $l_{max}$, the radius $r_{fms}$ of the cluster of atoms used in the full-multiple-scattering calculations, and a small imaginary part $\gamma$ added to the energy of the Green's functions; the value of $\gamma$ is set to $0.1$ eV for all calculations except for the electron gas where it is set to $0.01$ eV. $R_{max}$ is treated as a convergence parameter for metallic systems and set to the Norman radius for   insulators.
By default, we use a dense logarithmic grid e.g.,  $r_{i,n}=\exp(-8.8 + 0.05 n)$, as in the Dirac-Fock atomic calculations in FEFF10.\cite{Descl}   Typically this amounts to about 200 points per cell for each frequency,  which is computationally manageable,
though a   sparser grid may be adequate in some cases.
For the calculations of the cumulant spectral function, broadening by a Voigt function with both Gaussian and Lorentzian half-widths of 0.25 eV is used for all systems. Finally, for calculations of the 2p XPS, several additional parameters were used. First,  a parameter is introduced to account for the spin-orbit splitting between the $2p_{1/2}$ and $2p_{3/2}$ contributions; second, a Gaussian broadening with width $\Gamma_{exp}$ was applied to the spectrum to  account for experimental broadening, and separate Lorentzian widths $\Gamma_{2p_{1/2}}$ and $\Gamma_{2p_{3/2}}$ to   account the different for core-hole lifetimes; third, a Shirley background\cite{shirleyBG} was calculated from, and then   added to the theoretical spectrum with a scaling  parameter $B$   to match the low energy limit of the experimental data.
 Below  we present results only for the local RSGF spherical approximation, which is found  to be semi-quantitative in comparison to experiment for the systems treated here.  A more complete treatment including longer-ranged contributions with the generalized RSMS formalism is reserved for the future.


\section{Results}

\subsection{Homogeneous electron gas}

\begin{figure}[ht]
\includegraphics[width=1.0\columnwidth]{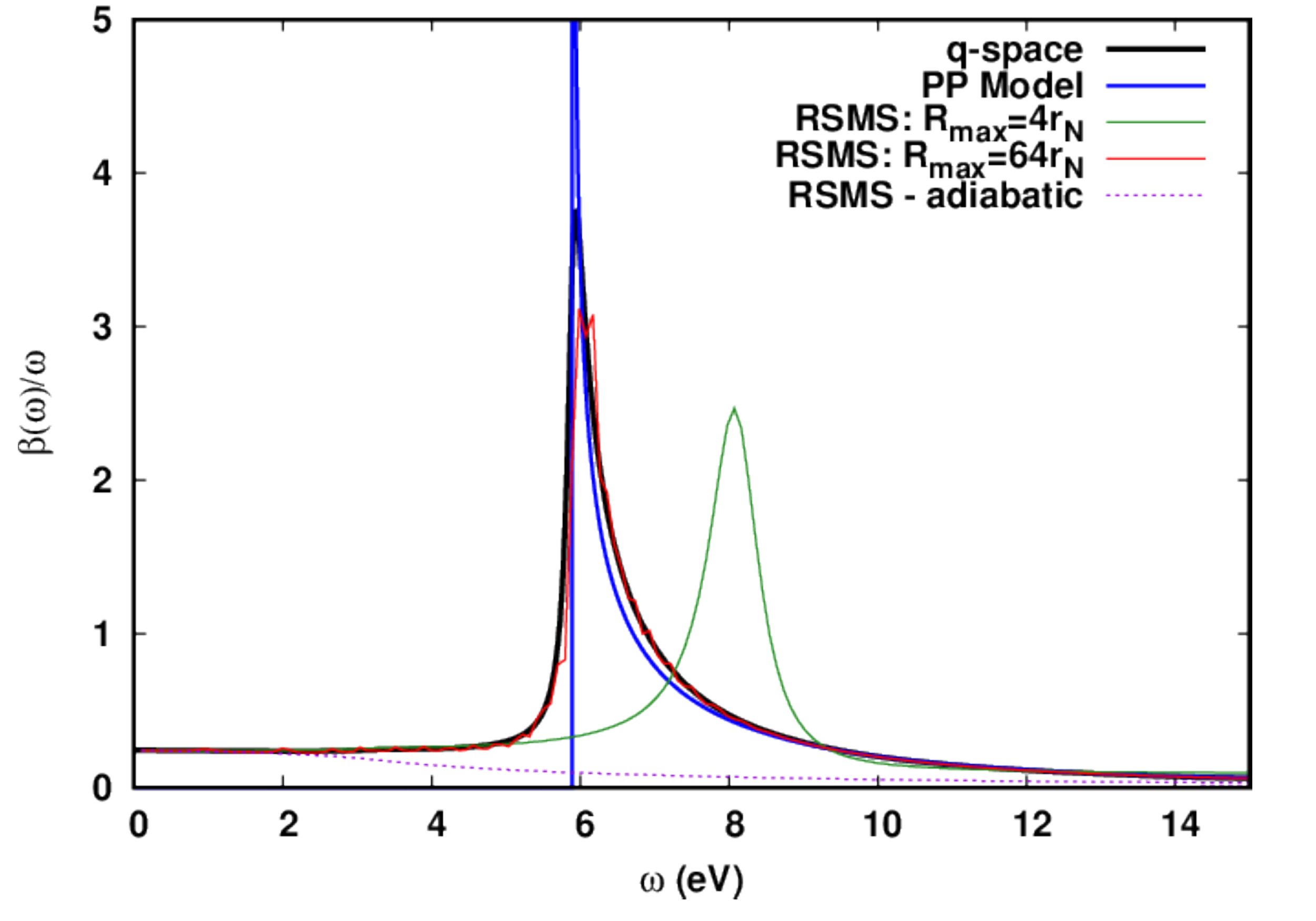}
\caption{Cumulant kernel $\beta(\omega)/\omega$ for the HEG at $r_s=4$    calculated using the RSGF approach compared to that based on the $q-$space formula in Eq.\ (\ref{beta-q}) and the plasmon-pole model.\cite{LundqvistII,hedin99rev} RSMS calculations are shown for $R_{max} = 4 r_N$ and $R_{max}=64\, r_N$, where the Norman radius  $r_N=1.4$ \AA\ was chosen as that for bcc 
sodium. Note that the low frequency, nearly constant behavior of $\beta(\omega)/\omega$, gives the edge singularity index $\alpha$. The dashed purple curve shows the adiabatic result which gives a good approximation for $\alpha$. 
    \label{fig:betaFEG}}
\end{figure}

\begin{figure}[ht]    
\includegraphics[width=1.0\columnwidth]{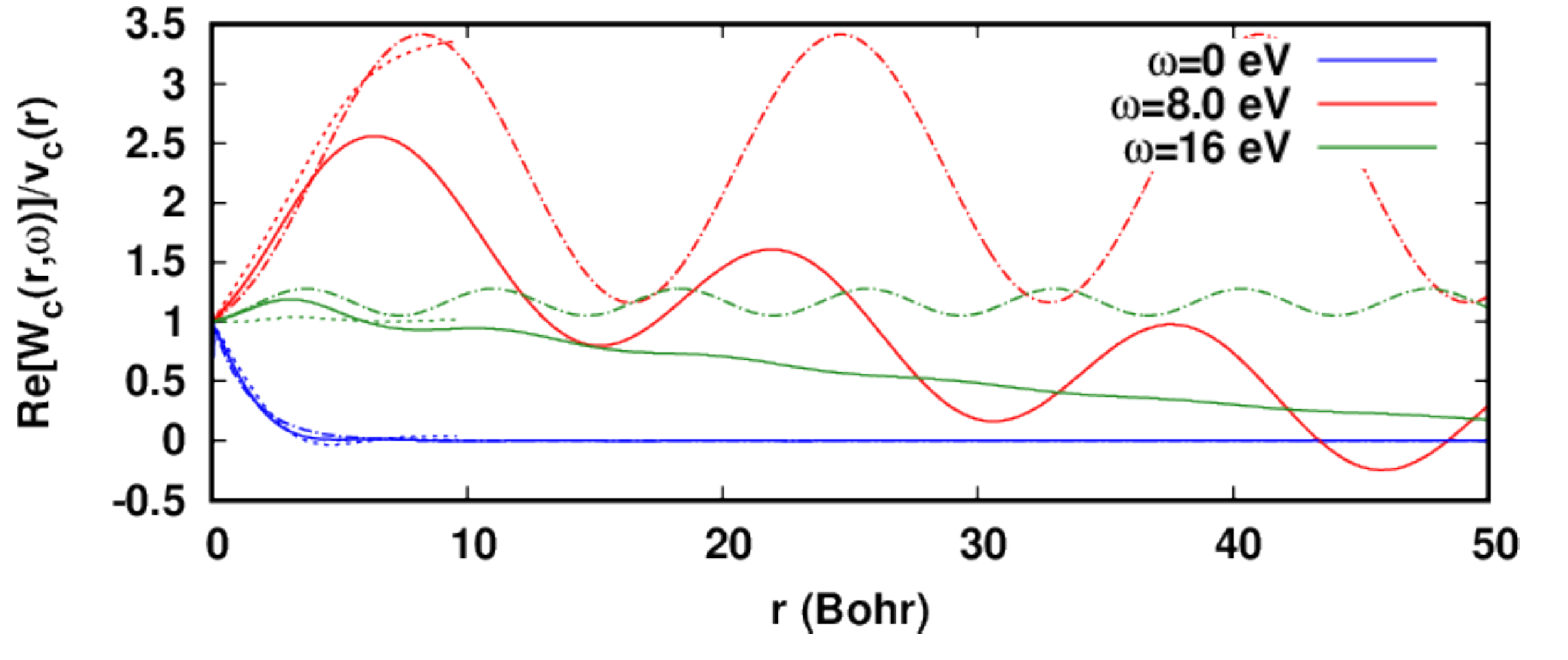}
\includegraphics[width=1.0\columnwidth]{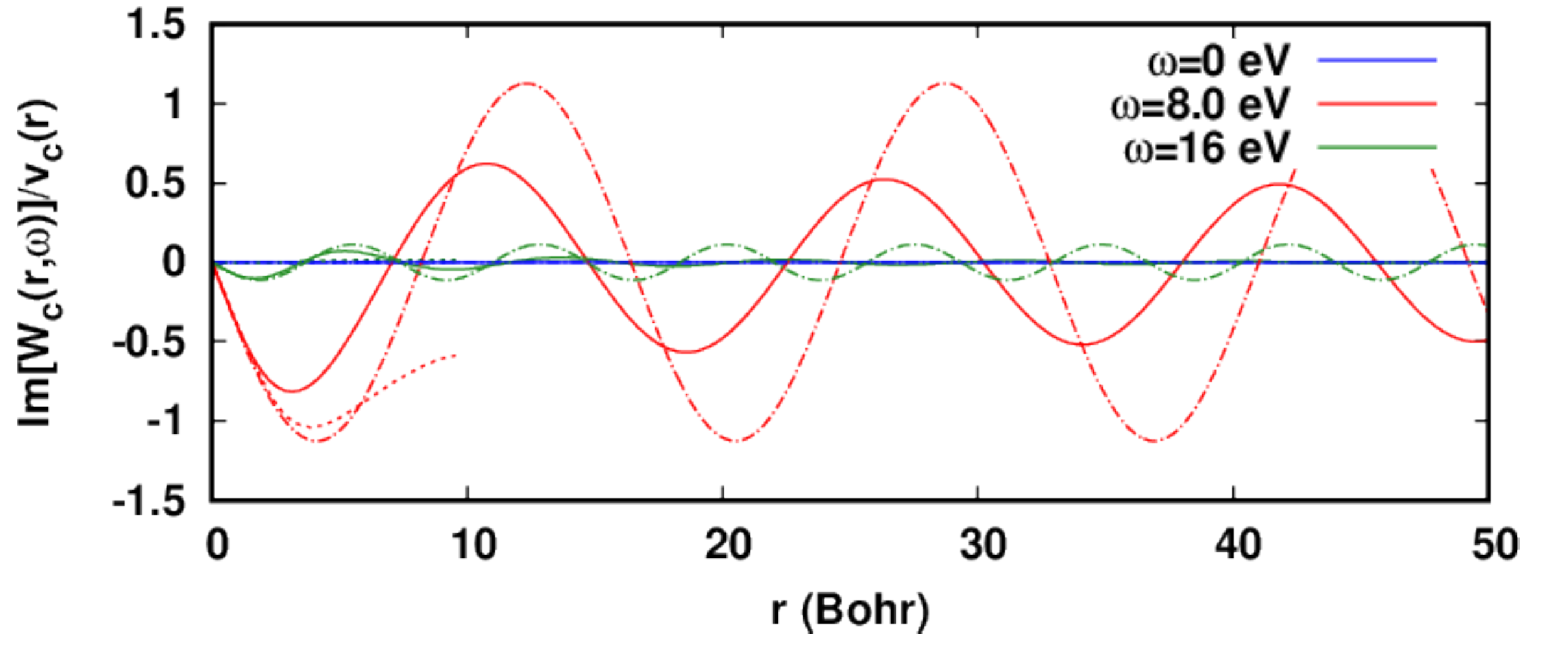}
    \caption{Screened core-hole potential from the RSGF approach (solid) relative to the bare core potential ${\rm Re}\, w_c(r,\omega)/V_c(r)$ (top) and ${\rm Im}\,
    w_c(r,\omega)/V_c(r)$ vs $r$ for the HEG at $r_s=4$ for $\omega=0$, $\omega=8$ eV $\approx \omega_p$  , and $\omega = 16 {\rm eV} >>\omega_p$. 
    Note that the behavior becomes strongly delocalized, oscillatory, and lossy near $\omega_p$. For comparison results from the RSRT approach (dashes) and the plasmon pole model (dot dashes) are also shown.
    \label{fig:wscrnFEG}}
\end{figure}

 As a first illustration and 
   proof of principle, we present results for the homogeneous electron gas (HEG).  An approximate treatment of the cumulant for this system was described by Langreth,\cite{langreth70}  and in more detail   by Lundqvist and Hedin \cite{LundqvistII,hedin99rev} for the plasmon-pole (PP) and RPA approximations.  
 For a deep core-level, the   core  charge density can be approximated by  a point charge,  so that the excitation spectrum $\beta(\omega)$ is given   by a one-dimensional integral in momentum space,
\begin{equation}
    \beta(\omega) = \frac{2}{\pi^2} \int d q\, {\rm Im}\, [\epsilon^{-1}(q,\omega)] \theta(\omega).
    \label{beta-q}
\end{equation} 
 This can be computed numerically within the RPA, or analytically with the plasmon pole model.\cite{suppmat} The results are shown in Fig.\ \ref{fig:betaFEG}, where $\beta(\omega)/\omega$ calculated with RSGF at several truncation radii $r_{max}$ are compared to the numerically exact $q-$space results and with the single plasmon-pole model.\cite{LundqvistII}
Consistent with the $q$-space formulation,
the behavior of $\beta(\omega)$ exhibits 
 a  linear part at low energy due to the particle-hole continuum, where
$-{\rm Im}\,\epsilon({\bf q},\omega)^{-1}\approx{\rm Im}\, \epsilon(q,\omega)/|\epsilon(q,0)|^2 $ and $\epsilon(q,0)$ is finite, as well as a sharply peaked structure 
above the plasmon onset $\omega_p \approx 6.0$ eV corresponding to the zeros of $\epsilon(q,\omega)$. The adiabatic approximation is also shown,  which compares well at low frequencies. 
Note that the RSGF formulation (red) performs  well at all energies except near the plasmon peak, where it loses some spectral weight. This is likely due to the truncations  of $R_{max}= 64\, r_N$ 
and $l_{max}=25$. The calculations also agree well with the PP model\cite{hedin99rev} above the plasmon frequency $\sim 6$ (eV); however the PP  approximation completely   ignores the low energy   contribution.\cite{suppmat} This low frequency tail in the HEG is dominated by the particle-hole contribution to the dielectric function and is linear in $\omega$. This behavior accounts for the Anderson edge singularity   in x-ray spectra and an asymmetric quasi-particle line-shape in XPS $J_k(\omega)\sim\omega^{\alpha-1}$.   For the HEG for $r_s=4$, we  obtain $\alpha=0.24$, which compares well to the experimental value $0.21\pm0.015$ for the 1s XPS of sodium, a nearly free electron system with $r_s \approx 4$. \cite{citrin1977} 
We also analyzed the extent to which the calculation of $\beta(\omega)$ can be approximated locally. The green curve in Fig.\ \ref{fig:betaFEG}  shows the result from RSGF calculations with $R_{max}=4r_{N}$ in the radial arguments of $\chi^0(r,r',\omega)$. While the linear portion is reproduced very well, the position of the plasmon peak is overestimated by $\sim 2$ (eV) or $30\%$. Although   difficult to see, the high energy tail is poorly represented and only corrected with much larger $r_{max}=64 r_N$.  
The RSGF spectrum approaches the $q-$space result 
with increasing $R_{max}$, but  does not  converge until $R_{max} =64 r_{N} \sim 177$ Bohr. 
This difference in convergence in different energy ranges reflects the fact that the sharp peak in $\beta(\omega)$ is largely defined by the sharp plasmon peak in the loss function, which is dominated by momentum transfer $q$ near zero. On the other hand, the linear behavior near $\omega=0$ and the long tail at high frequency are due to the particle-hole continuum and the dispersion of the plasmon respectively. These     depend on higher values of momentum transfer  which results in   faster convergence with $R_{max}$. 
Fig.\ \ref{fig:wscrnFEG} shows the screened core-hole potential
$w_c(r,\omega)$ (also for $r_s=4$) scaled for convenience by the bare potential $V_c(r)$ vs $r$ for  frequencies from $0$ eV  to well above the $\omega_p$. The RSGF result (solid) are compared with the results of the plasmon pole (dot dashes) and those from the RSRT formulation (dashes). 
As expected physically, there is   little screening at small $r<<r_s$ for all frequencies due to the slow build-up of the induced density response. For $r>r_s$ the zero frequency curve (red)   exhibits extremely efficient Yukawa-like screening, while the behavior at high frequencies is weakly screened, with
$w_c(r,\omega)$ remaining finite at large $r$.
Near the plasma-frequency $\omega_p\approx 6$ eV $w_c(r,\omega)$ oscillates sinusoidally,
corresponding to   delocalized charge density 
fluctuations with substantial contributions to ${\rm Im\,} w(r,\omega)$. The oscillatory behavior matches reasonably well with that of the plasmon pole model, although there is more damping and larger screening in the high frequency curves calculated with RSGF, and the oscillations are slightly phase shifted. Some of these differences are expected due to differences in the plasmon dispersion and the lack of particle-hole continuum states in the plasmon pole model. The majority of the damping seen in the RSGF results comes from the truncation of the real-space grid, although this damping does not affect the calculated values close to the origin, which determine $\beta(\omega)$. 

\subsection{Nearly free-electron system}

\begin{figure}[t]
    \includegraphics[width=1.0\columnwidth]{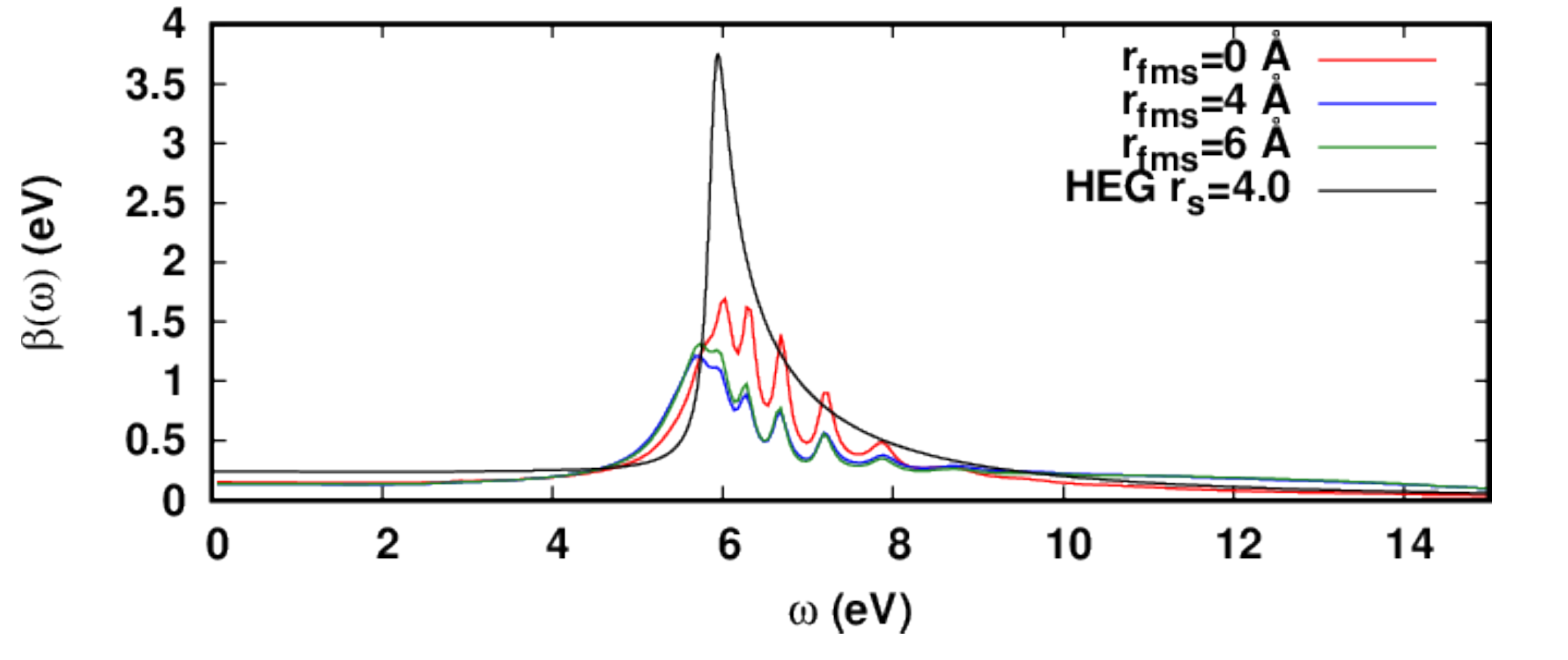}
    \includegraphics[width=1.0\columnwidth]{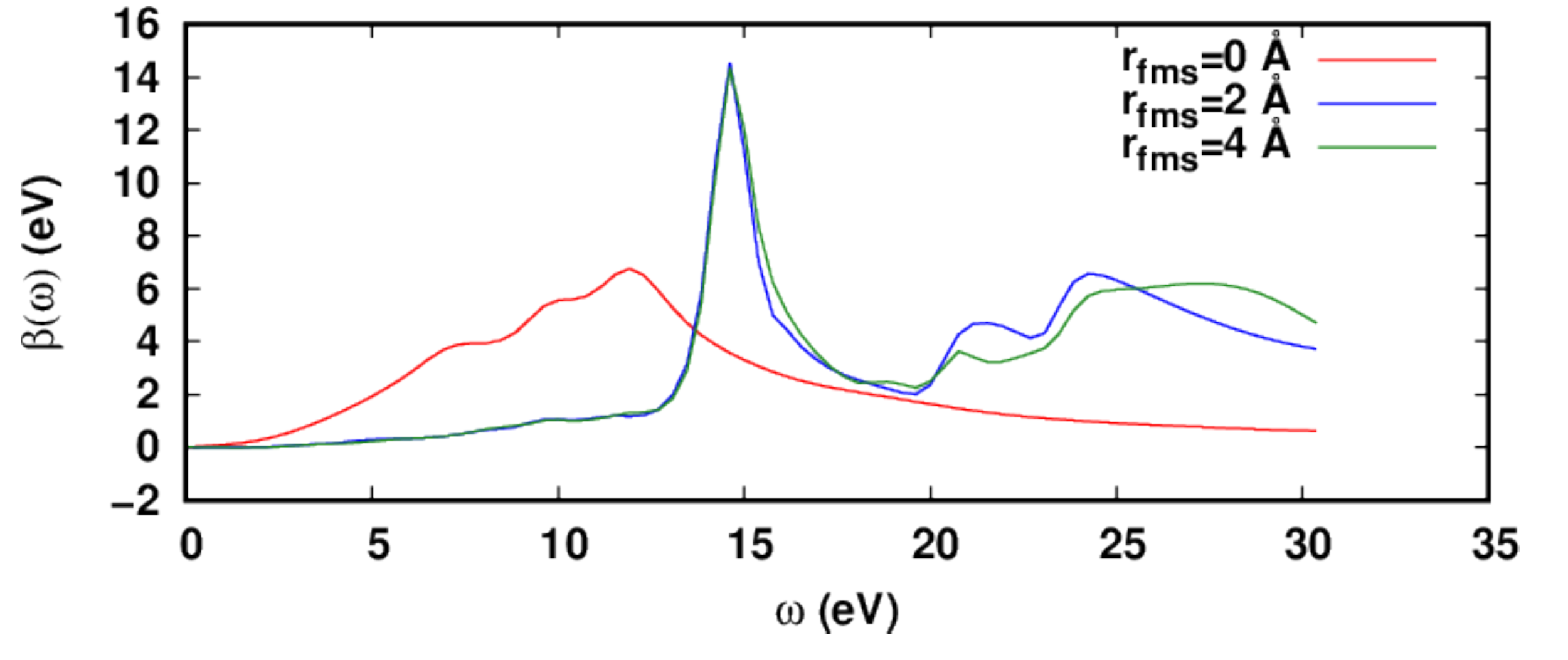}  
    \caption{RSGF calculation of the cumulant kernel $\beta(\omega)$ for Na (top) and TiO$_2$ (bottom,
    with varying FMS cluster size $r_{fms}$.
    Note that scattering from the nearest-neighbor atoms has a drastic effect in TiO$_2$ but is much smaller for Na. Both spectra show minimal differences with theinclusion of atoms beyond the nearest-neighbors.
    \label{fig:betaFMS}}
\end{figure}

As an example of the RSGF approach that can  be compared with the HEG, we present calculations for  bcc Na, a nearly free electron (NFE) system with $r_s \approx 4$.
This is illustrated in Fig.\ \ref{fig:betaFMS} for a range of $r_{fms}$ and for large $R_{max}$. The  behavior of $\beta(\omega)$ is clearly linear below $\omega_p$ and comparable to that of the HEG at high energies as well.  Moreover, the behavior near the plasmon peak also has long range oscillatory contributions.   For such NFE systems, scattering from near-near neighbors appears to be less important than a large $R_{max}$ alone, as the results converge reasonably well by setting $r_{fms}=0$.  The oscillatory behavior near the peak reflects the discrete nature of  plasmon excitations within a sphere of finite $R_{max}$.

 \subsection{Transition metal compounds}
  
\begin{figure}[ht]
  \includegraphics[width=1.0\columnwidth]{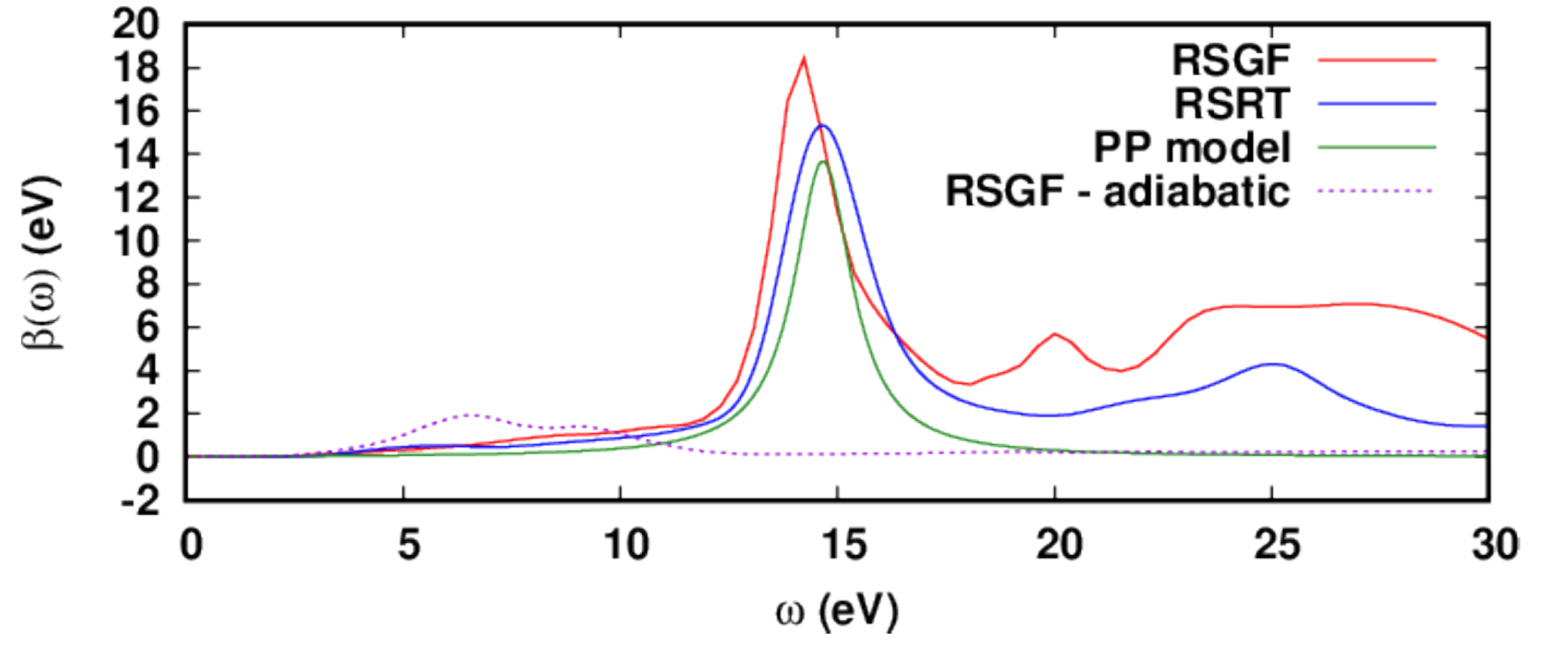}
    \includegraphics[width=1.0\columnwidth]{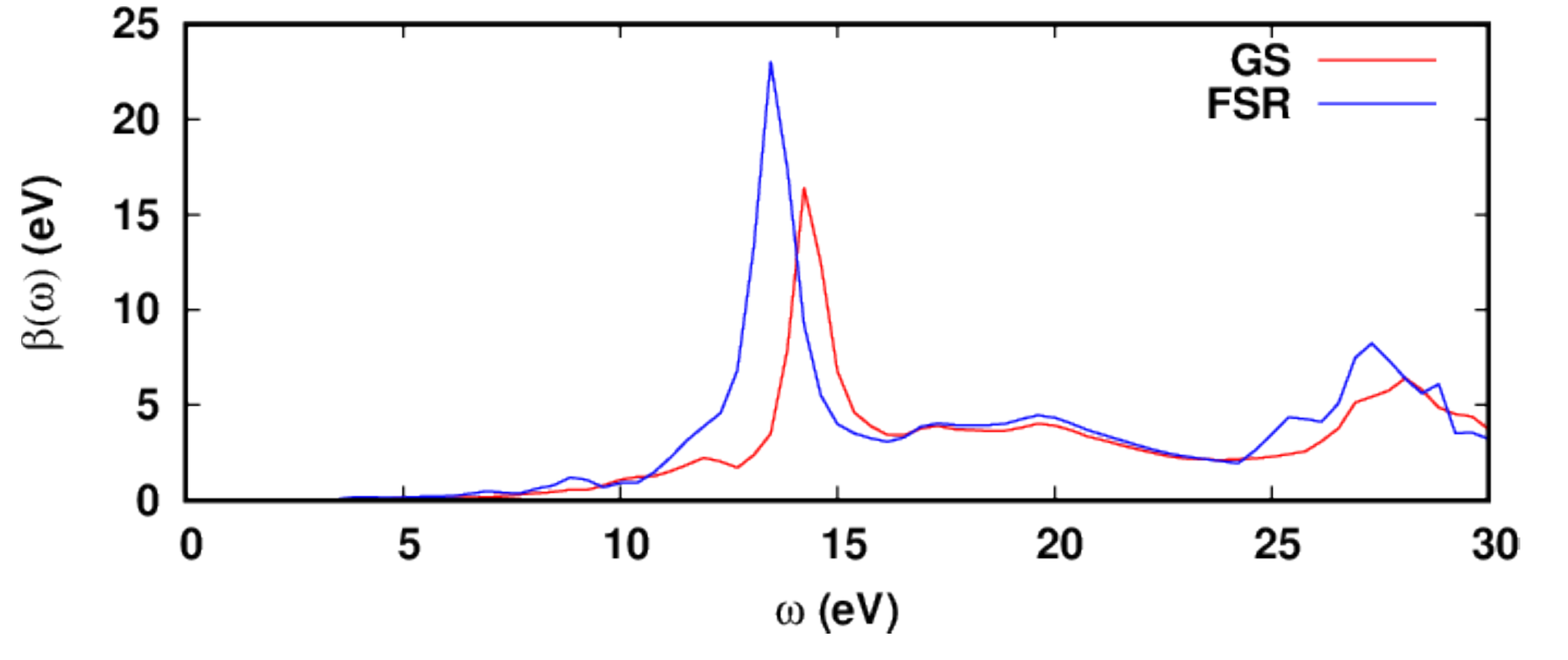}
    \caption{Cumulant kernel $\beta(\omega)$ for TiO$_2$ (Top) calculated with the RSGF approach (red), and for comparison the RSRT (blue),  the fluctuation potential approach with a single pole (PP), and the adiabatic limit (dashed). The lower figure for ScF$_3$ (Bottom) compares results calculated with ground state electronic structure (GS), and with the final state rule (FSR). The core-hole shifts the peak to lower energy by $\sim 0.8$ eV, in better agreement with experiment.
    \label{fig:beta-rutile}}
\end{figure}

Next we consider several early transition metal compounds,
 which are typically moderately correlated.   We first
present results for TiO$_2$ (rutile), which was previously studied  with the RSRT approach.\cite{Kas2015}
The strong peak  in $\beta(\omega)$ for TiO$_2$ is shown in Figs.\ \ref{fig:betaFMS} (bottom) and \ref{fig:beta-rutile} (top). Unlike the behavior of Na, the spectrum is largely independent of $R_{max}$, while $r_{fms}$ has a drastic effect on the position and intensity of the main peak in $\beta(\omega)$ as seen in Fig.\ \ref{fig:betaFMS}. 
The independence of the spectrum with respect to $R_{max}$ suggests a localized nature of the excitations in TiO$_2$. The main peak is neither directly related to peaks in $\chi^0(\omega)$ nor in ${\rm Im}\, \epsilon(\omega)$. Instead the peak is consistent with zero-crossings of ${\rm Re}\, \epsilon(\omega)$. For TiO$_2$ we have found that there is a single dominant eigenvalue of the dielectric matrix at $\omega_{p}$, although the lack of more than one crossing may be due to the limited spacial extent of our calculations. 
Consequently a fluctuation potential treatment [cf. Eq.\ (22) and (23)] is appropriate. This PP model yields a Lorentzian  shape for 
$\beta(\omega)$ and ${\rm Im}\, w_c(r,\omega)$, as seen in Fig.\ \ref{fig:beta-rutile} (top, green)  as well as a finite range of  $w_c(r,\omega)$  varying inversely with $\gamma_p$. 
The  cumulant kernel $\beta(\omega)$ 
calculated with  
Eq.\ (\ref{betaspherical}) and $R_{max}=r_{N}$ is shown in Fig.\ \ref{fig:beta-rutile} (top), along with a comparison to   the RSRT method.  
\begin{figure}[ht]  
 \includegraphics[width=1.0\columnwidth]{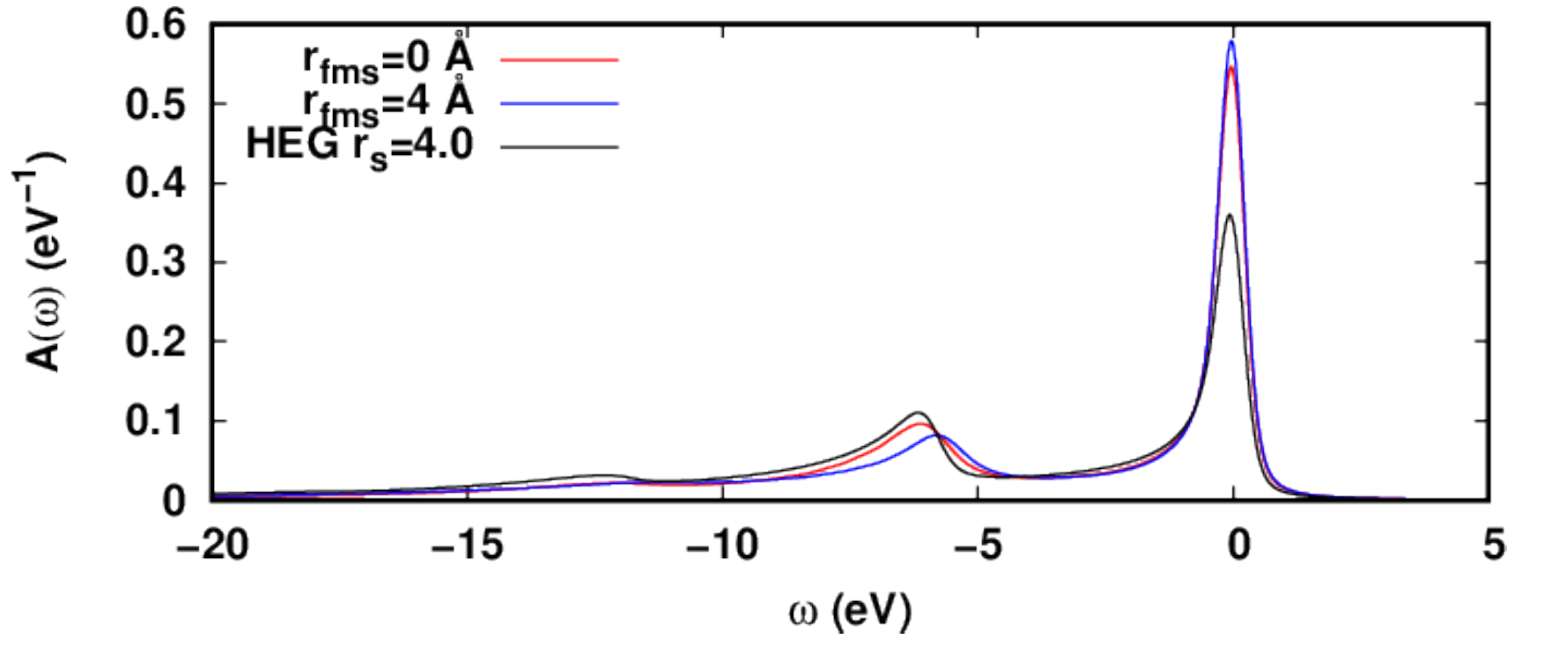}
\includegraphics[width=1.0\columnwidth]{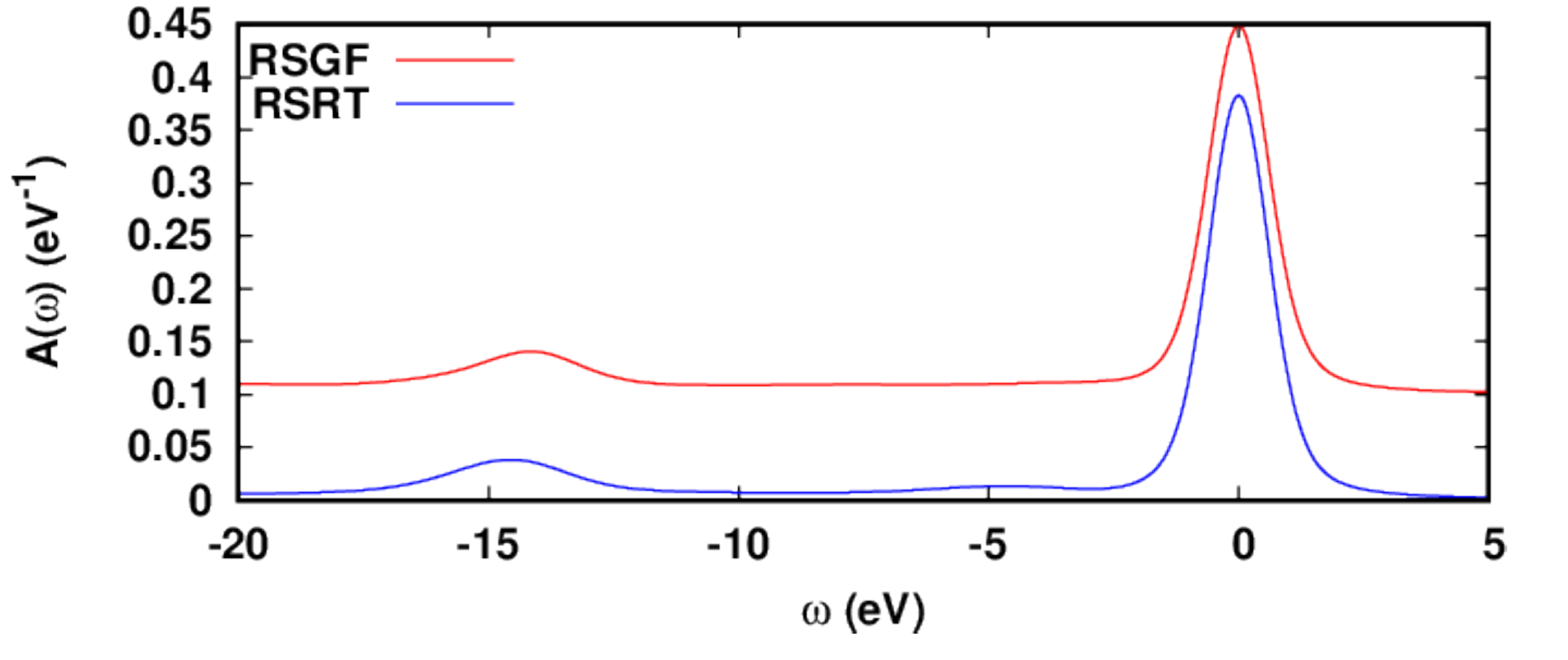}
   \includegraphics[width=1.0\columnwidth]{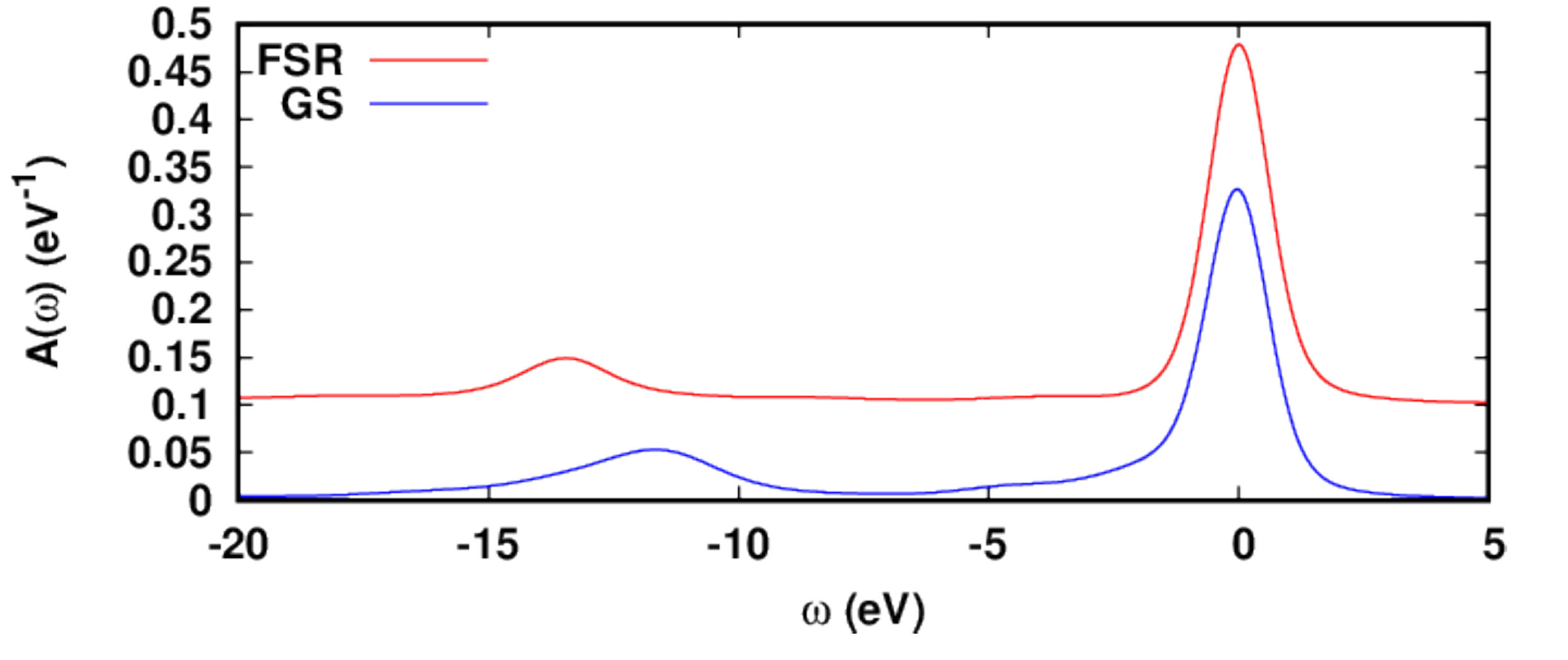}
    \caption{Spectral functions $A(\omega)$for 
   Na (Top) compared with the HEG with different $r_{fms}$.  The middle figure compares the RSGF and RSRT\cite{Kas2015} approaches for TiO$_2$,  and the bottom figure compares the spectral function of ScF$_3$ calculated with (FSR) and without (GS) the final state rule core-hole.  
    \label{fig:spectralfnRutile}
    }
\end{figure}

   In contrast to the HEG, the adiabatic approximation (top, purple) does not reflect the spectrum of $\beta(\omega)$, even at low frequencies, and is in error by $\sim 50\%$ by $\approx 2.5$ eV, which is about where the first excitation feature occurs. Instead, almost all structure in $\beta(\omega)$, or at least that above $\approx 7$ eV, reflects that of $|W_c(\omega)|^2$.
   Below $7$ eV, the spectrum has little structure. 
The prominant peak  at about 15 eV  stems from the single pole of $\epsilon^{-1}(\omega)$ where $\lambda_p(\omega)$ and hence ${\rm Re}\, \epsilon (\omega) $ crosses zero. 
 The low energy structure is similar for both RSGF and RSRT methods, and the agreement of the peak near 15 eV is semi-quantitative. Finally, at higher energies there is appreciable difference between the RSGF and RSRT results, which could be due to the limited basis set and lack of continuum states in the RSRT results, which rely on local basis functions, or the local approximation of the RGSF method.
 Note however, that the structure above 15 eV does not contribute significantly to the spectral function, due to the factor of $1/\omega^2$ in the cumulant $C_c(t)$ in Eq.\ (\ref{eq:landau}). 
Results for ScF$_3$ (bottom) are similar, with a large peak at about $13$ to $14$ eV depending on whether the calculation was performed with (FSR), or without (GS) a core-hole. The inclusion of the core-hole red-shifts the spectrum, and increases the main peak intensity, similar to the effects seen in x-ray absorption near-edge spectra. This excitonic shift improves the agreement with experiment, which shows a peak at about $12$ eV (see Fig.\ \ref{fig:ScXPS}).  The spectral functions, which are closely related to XPS, are shown in Fig.\ \ref{fig:spectralfnRutile} for Na, TiO$_2$ and ScF$_3$, calculated using Eq.\ (\ref{spectralfn}).  The spectral function of Na is compared with that of the homogeneous electron gas, and is shown for several values of $r_{fms}$, while that of TiO$_2$ (middle) is compared to the result from RSRT, and shows good agreement. Finally, the spectral function of ScF$_3$ is shown as calculated with (FSR) and without (GS) a final-state-rule core-hole. In these curves, we see a large main peak at high energy, corresponding to the quasiparticle peak, and satellite peaks at lower energy, which reflect the behavior of $\beta(\omega)/\omega^2$. The quasiparticle peak in the spectral function of Na shows an appreciable asymmetry, corresponding to the edge singularity, and originating from the linear behavior in $\beta(\omega)$ at low frequency.
\begin{figure}[ht]
    \includegraphics[width=1.0\columnwidth]{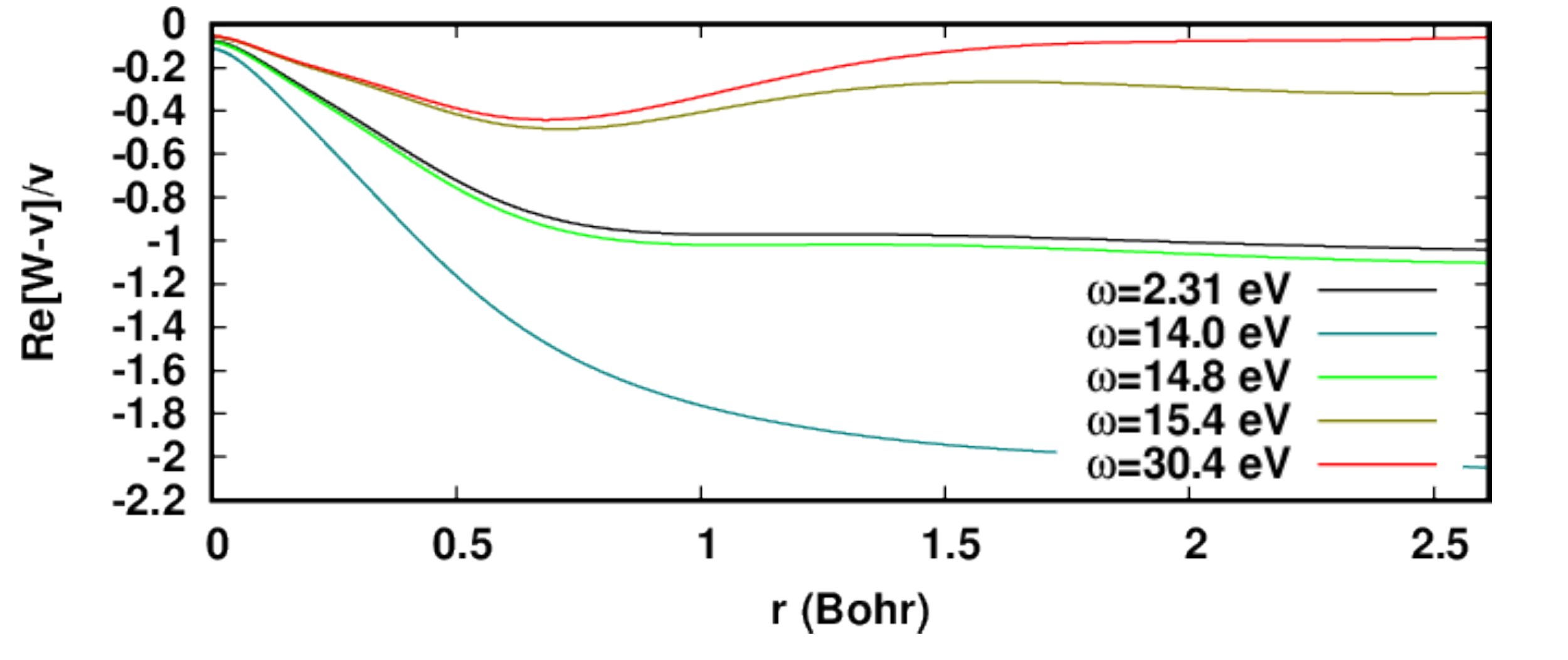}
    \includegraphics[width=1.0\columnwidth]{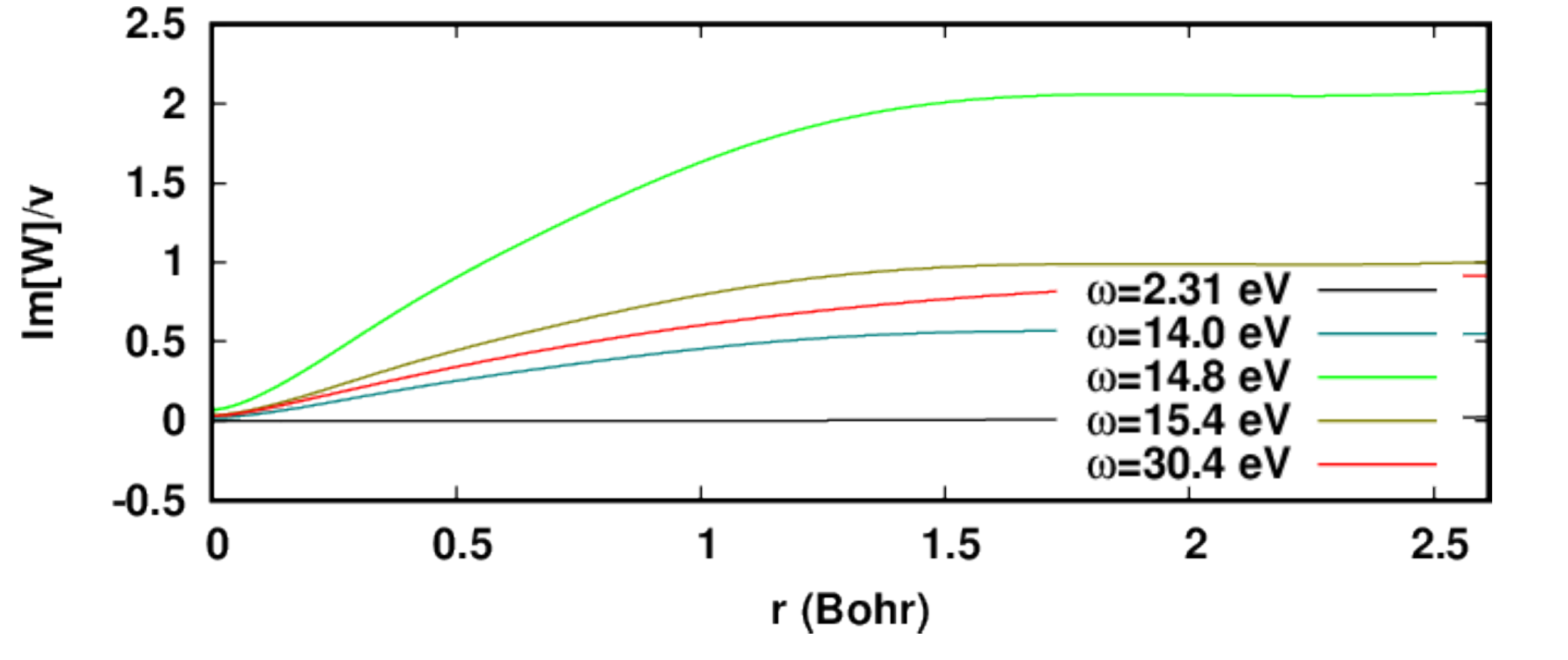}
    \caption{Screening fraction $f(r,\omega)=[w_c(r,\omega)-V_c(r,\omega)]/V_c(r)$ relating the dynamically screened $w_c(r,\omega)$ and bare $V_c(r)\approx 1/r$ core-hole potentials for the Ti core-hole in TiO$_2$ at a few selected energies. Note that the screening is local and strong at low (black), weak at high (red), and becomes lossy and delocalized near $\omega_{CT}$ with roughly a Lorentzian line-shape. The behavior at large $r$ is only semi-quantitative due to the truncation of $r_{max}$.
    }
    \label{fig:wscrn}
\end{figure}

The dynamical screening fraction of the core-hole potential which here is defined as $f(r,\omega) = [w_c(r,\omega)-V_c(r)]/V_c(r)$,    calculated from Eq.\ (\ref{eq:wcr}) with  the local RSGF cumulant is shown for Rutile TiO$_2$  in Fig.\ \ref{fig:wscrn} for a few selected energies. The bare core-hole potential $V_c(r)$ is  that for the $1s$ core-level of Ti and is nearly Coulombic for a point charge, i.e., $V_c(r) \approx 1/r$ beyond the Ti $1s$ radius $r_{1s}\approx 1/22$ Bohr. The screening fraction $f(r,\omega)$ is  small near $r=0$ (i.e., little screening), and fairly strong  $f(\omega) \approx -1$ beyond the Thomas-Fermi screening length $r_0$. However, near the peak at $\omega_p\approx 14.8$ eV, abrupt changes occur, with the system becoming over-screened ($f(\omega)$ reaching about $-2$ at $14$ eV) or   under-screened  ($f(\omega)<0.25 $) above $15.4$ eV. 
The imaginary part of $W_c$ largely reflects the structure in $\beta(\omega)$, starting at $0$ for $\omega=0$, and rising to a peak at $\omega_p\approx 14.8$ eV, then rapidly subsiding, although it is still appreciable at high energy. 
Note that the behavior of $W_c(r,0)$ is metallic,  quickly going to zero beyond the Thomas-Fermi screening radius, which is likely due to both the finite cluster size in the calculation of the Green's functions, which produce  only a pseudo-gap in the material, and the neglect of off-diagonal elements in the interaction kernel $K$ and response function $\chi^0$. However, this unphysical behavior does not seem to
affect $\beta(\omega)$ appreciably. 
We also note that the satellites in the TiO$_2$ spectral function correspond to the $\approx 13$ eV peak seen in experimental XPS, which have been interpreted both as charge-transfer excitations,\cite{Kas2015,woicik1,woicik2,bocquet1996} as well as plasmons.\cite{vast2002} Although  distinction between plasmon vs charge transfer excitations may be a question of semantics,\cite{miyakawa1968,zupanovic1997,gunn1979}   the behavior of the excitation in TiO$_2$ and the other transition metal compounds is rather different from the plasmons found in FEG-like materials, where the lattice has little effect on the plasmon energy. The apparently localized nature of the excitations in TiO$_2$ is consistent with the definition of charge transfer excitations. Nevertheless, for the systems investigated here, the excitations at $\omega_p$ corresponds to a zero crossing of the dielectric matrix. 

\begin{figure}[ht]  
\includegraphics[width=1.0\columnwidth]{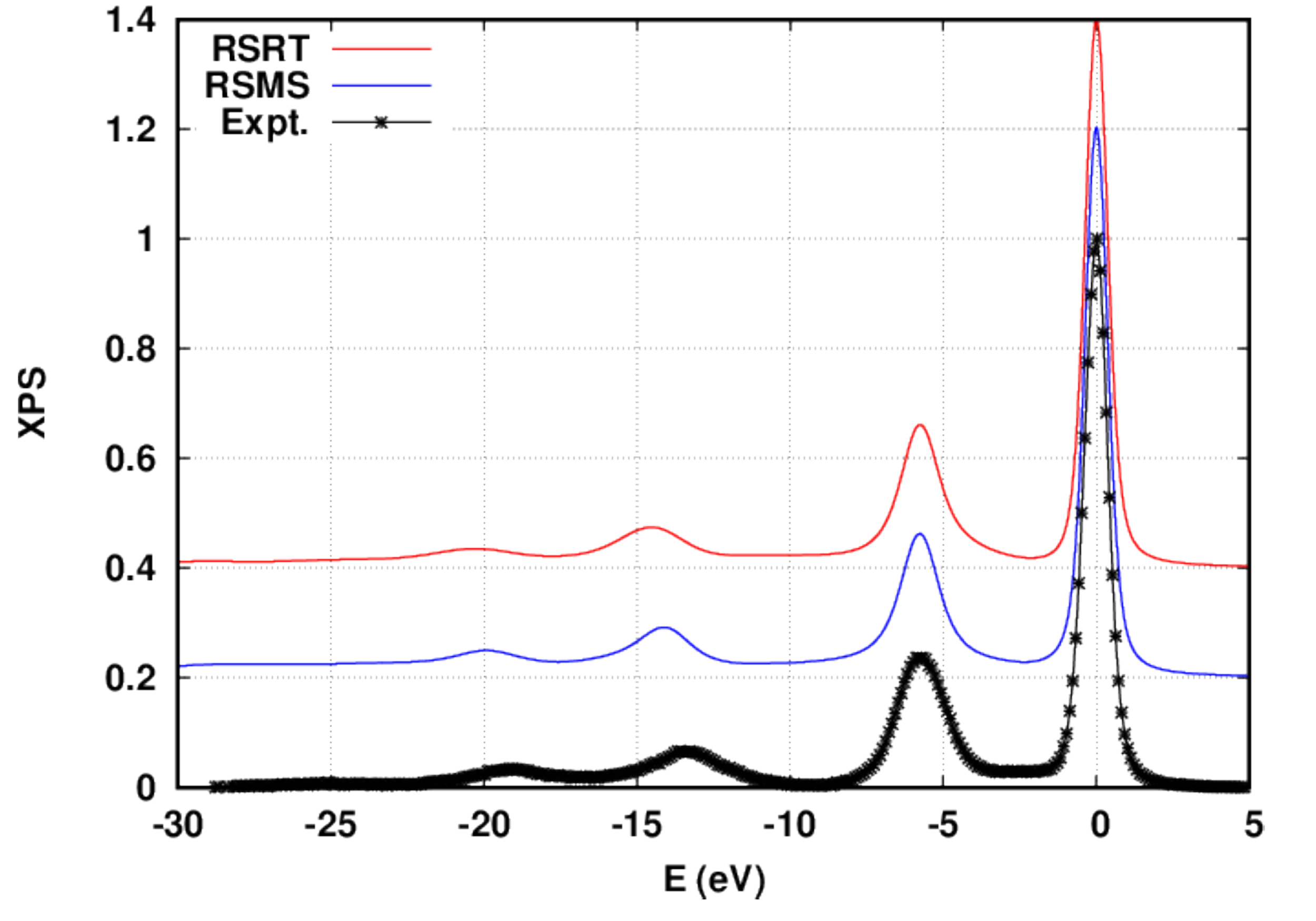}
    \caption{Comparison of the RSGF Ti 2p spectral function with the RSRT approach  and XPS experiment.\cite{deboer}  
  \label{fig:TiXPSvsExpt}
  }
\end{figure}

\begin{figure}[ht]  
\includegraphics[width=1.0\columnwidth]{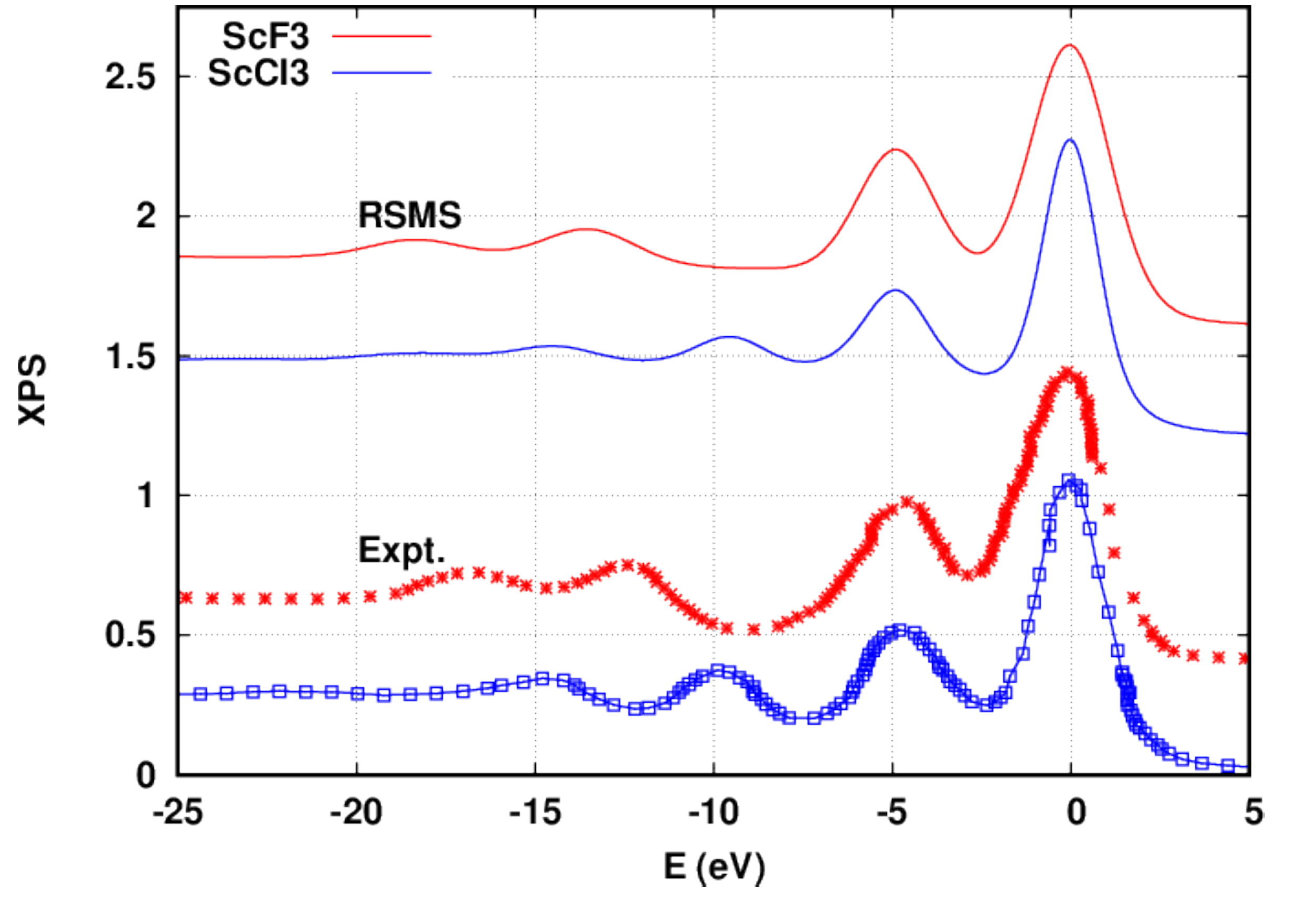}
    \caption{Comparison of the Sc 2p XPS of
    ScF$_3$ (red) and ScCl$_3$ (blue) calculated with the RSGF approach and with experiment (dots). The experimental data was digitally reproduced from Ref.\ \onlinecite{deboer}
  \label{fig:ScXPS}
  }
\end{figure}

Finally, we compare calculated and experimental metal $2p$ XPS of TiO$_2$, ScF$_3$, and ScCl$_3$ in Figs.\ \ref{fig:TiXPSvsExpt} and \ref{fig:ScXPS}, which show the spectra as a function of energy relative to the $2p_{3/2}$ main peak. The main  $2p_{3/2}$ peak is centered at $E=0$ eV, and the first peak below that is the  $2p_{1/2}$ main peak, seen at $\approx-6$ eV in TiO$_2$, and $\approx -5$ eV in the Sc halides. Below the  $2p_{1/2}$ peaks, two satellites can be seen, corresponding to a many-body excitation beyond the main  $2p_{3/2}$ or $2p_{1/2}$ peak, lowering the photoelectron energy by the energy of the excitation $\Delta E$.
Experimentally, the satellite energy is the largest in TiO$_2$ at $\Delta E\approx 13.4$ eV, followed by ScF$_3$ with $\Delta E\approx 12.3$, then ScCl$_3$ at $\Delta E\approx10$ eV. This trend is reproduced by the calculations, although the satellite energies are slightly overestimated for TiO$_2$ and ScF$_3$, and slightly underestimated for ScCl$_3$, with $\Delta E\approx14.1$ eV, $13.4$ eV, and $9.5$ eV respectively. The intensities of the satellites are also slightly underestimated by the local RSMS theory as well as that of the RSRT (shown for TiO$_2$). The satellite energies are inversely correlated with the metal ligand bond-length in the material, with the bond-lengths at $1.95$ \AA\  for TiO$_2$, $2.03$ \AA\  for ScF$_3$, and $2.51$ \AA\ for ScCl$_3$. This is reminiscent of XAS, where an expansion of the bond contracts the oscillatory fine-structure in the spectrum. This likely explains the trend between ScF$_3$ and ScCl$_3$, although the differing oxidation state of Ti in TiO$_2$ may also play a role in the satellite energy. Overall, the agreement between the local RSMS theory and experiment is   remarkable given the level of approximations made.

\section{Summary and Conclusions}

  We have developed an {\it ab initio} real-space Green's function approach for calculations of    intrinsic inelastic
  losses in x-ray spectra. These losses show up e.g., as satellites in XPS and reduced fine-structure amplitudes in XAS, but are often  neglected in conventional calculations. The approach is based on the cumulant Green's function formalism and a generalization of the Langreth cumulant to inhomogeneous systems
  analogous to that in the RSRT, TDDFT approach. The formalism of the cumulant kernel $\beta(\omega)$ is   analogous to 
 x-ray and optical absorption spectra $\mu(\omega)$, except that the (dipole) transition operator is  replaced
by the core-hole potential $V_c({\bf r})$, with monopole selection rules. Within the RPA, three equivalent expressions for $\beta(\omega)$ are derived, and the link to the GW self-energy is discussed. 
  A generalized real-space multiple scattering (RSMS)
formalism with a discretized site-coordinate basis is introduced to carry out the calculations with standard linear algebra. The computational effort further simplifies with spherical symmetry of the screened core-hole potential and response function.
The calculations can then be carried
out on a one-dimensional radial grid, similar to that in conventional atomic calculations and  the RSGF approach for XAS.  Solid-state effects from the extended system are included in terms of RSGF calculations of back scattering contributions from atoms beyond the absorption site. 
The behavior of both $W_c({\bf r},\omega)$ and $\beta(\omega)$ reflect the analytic structure of the inverse dielectric matrix and the loss function, where  the peaks in $\beta(\omega)$ arise from the  zeros of  ${\rm Re}\, \epsilon(\omega)$. This behavior is consistent with delocalized quasi-boson excitations, which have been interpreted both as plasmons and charge-transfer excitations.
 Moreover, the low energy  $\omega\rightarrow 0$ limit, $\beta(\omega) \approx \alpha\omega$ in metallic systems, is consistent with the Anderson edge singularity exponent. 
Our local RSGF cumulant approximation yields results in semi-quantitative agreement with the complementary RSRT approach, as well as with experimental XPS, which validates the local approximations.  The discrepancy between RSGF and RSRT results is likely due to the neglect of the long range behavior in the bare response function and in the interaction kernel. This behavior is often not well captured by the truncation of the real-space calculations to small $R_{max}$.  Although this is one of the main limitations of the local model, corrections can be added using the more general RSMS formalism with a larger site-radial coordinate basis. Nevertheless, in transition metal oxides like TiO$_2$, ScF$_3$, and ScCl$_3$, the convergence is found to be fairly rapid, yielding reasonable accuracy compared with the RSRT approach. 
In conclusion, the RSGF approach illustrates  the nature and localization of the dynamically screened core-hole $W_c(r,\omega)$ and the cumulant kernel $\beta(\omega)$.
The approach  permits {\it ab initio}  treatments of  the spectral function
$A_c(\omega)$ and the quasi-particle XAS $\mu_1(\omega)$ in the convolution 
in Eq.\ (\ref{xasconv}), which can be calculated in parallel with the same RSGF formalism. The approach thereby permits a unified treatment of XPS and XAS that builds in intrinsic losses.    The method can also be applied {\it ex post facto} to include  intrinsic losses in other calculations of x-ray spectra. 
Many extensions are possible. For example,   corrections to the dielectric matrix from next neighbors may be desirable to improve convergence 
with respect to $R_{max}$.\cite{prange_opcons,tddft-bse} 
Although we focused on result within the RPA, generalization to a real TDDFT interaction kernel $f_{xc}$ poses no formal or numerical difficulties.
Our generalized RSMS formulation opens the possibility of RSMS approaches to optical spectra beyond the RPA, e.g., within the TDDFT or the Bethe-Salpeter equation formalism. Finally, contributions from extrinsic 
losses and interference terms can 
be relevant.\cite{campbell,cudazzo,Kas2015}
Further developments along these lines are reserved for the future.

\noindent {Acknowledgments} --
We thank T. Fujikawa, L. Reining, E. Shirley and J. Woicik for comments, and S. Adkins and S. Thompson for encouragement.  This work is supported by the
Theory Institute for Materials and Energy Spectroscopies (TIMES) at SLAC,
which is  funded by the U.S. DOE, Office of Basic Energy Sciences, Division of Materials Sciences and Engineering,
under contract DE AC02-76SF0051.


\begin{thebibliography}{46}
\expandafter\ifx\csname natexlab\endcsname\relax\def\natexlab#1{#1}\fi
\expandafter\ifx\csname bibnamefont\endcsname\relax
  \def\bibnamefont#1{#1}\fi
\expandafter\ifx\csname bibfnamefont\endcsname\relax
  \def\bibfnamefont#1{#1}\fi
\expandafter\ifx\csname citenamefont\endcsname\relax
  \def\citenamefont#1{#1}\fi
\expandafter\ifx\csname url\endcsname\relax
  \def\url#1{\texttt{#1}}\fi
\expandafter\ifx\csname urlprefix\endcsname\relax\def\urlprefix{URL }\fi
\providecommand{\bibinfo}[2]{#2}
\providecommand{\eprint}[2][]{\url{#2}}

\bibitem[{\citenamefont{Rehr and Albers}(1990)}]{rehralb}
\bibinfo{author}{\bibfnamefont{J.~J.} \bibnamefont{Rehr}} \bibnamefont{and}
  \bibinfo{author}{\bibfnamefont{R.~C.} \bibnamefont{Albers}},
  \bibinfo{journal}{Phys. Rev. B} \textbf{\bibinfo{volume}{41}},
  \bibinfo{pages}{8139} (\bibinfo{year}{1990}).

\bibitem[{\citenamefont{Zhou et~al.}(2020)\citenamefont{Zhou, Reining,
  Nicolaou, Bendounan, Ruotsalainen, Vanzini, Kas, Rehr, Muntwiler, Strocov
  et~al.}}]{zhou-pnas}
\bibinfo{author}{\bibfnamefont{J.~S.} \bibnamefont{Zhou}},
  \bibinfo{author}{\bibfnamefont{L.}~\bibnamefont{Reining}},
  \bibinfo{author}{\bibfnamefont{A.}~\bibnamefont{Nicolaou}},
  \bibinfo{author}{\bibfnamefont{A.}~\bibnamefont{Bendounan}},
  \bibinfo{author}{\bibfnamefont{K.}~\bibnamefont{Ruotsalainen}},
  \bibinfo{author}{\bibfnamefont{M.}~\bibnamefont{Vanzini}},
  \bibinfo{author}{\bibfnamefont{J.}~\bibnamefont{Kas}},
  \bibinfo{author}{\bibfnamefont{J.}~\bibnamefont{Rehr}},
  \bibinfo{author}{\bibfnamefont{M.}~\bibnamefont{Muntwiler}},
  \bibinfo{author}{\bibfnamefont{V.~N.} \bibnamefont{Strocov}},
  \bibnamefont{et~al.}, \bibinfo{journal}{Proc. Nat. Acad. Sci.}
  \textbf{\bibinfo{volume}{117}}, \bibinfo{pages}{28596}
  (\bibinfo{year}{2020}).

\bibitem[{\citenamefont{Langreth}(1970)}]{langreth70}
\bibinfo{author}{\bibfnamefont{D.}~\bibnamefont{Langreth}},
  \bibinfo{journal}{Phys. Rev. B} \textbf{\bibinfo{volume}{1}},
  \bibinfo{pages}{471} (\bibinfo{year}{1970}).

\bibitem[{\citenamefont{Bardyszewski and Hedin}(1985)}]{hedin-bard}
\bibinfo{author}{\bibfnamefont{W.}~\bibnamefont{Bardyszewski}}
  \bibnamefont{and} \bibinfo{author}{\bibfnamefont{L.}~\bibnamefont{Hedin}},
  \bibinfo{journal}{Physica Scripta} \textbf{\bibinfo{volume}{32}},
  \bibinfo{pages}{439} (\bibinfo{year}{1985}).

\bibitem[{\citenamefont{Hedin}(1999)}]{hedin99rev}
\bibinfo{author}{\bibfnamefont{L.}~\bibnamefont{Hedin}},
  \bibinfo{journal}{Journal of Physics: Condensed Matter}
  \textbf{\bibinfo{volume}{11}}, \bibinfo{pages}{R489} (\bibinfo{year}{1999}).

\bibitem[{\citenamefont{Campbell et~al.}(2002)\citenamefont{Campbell, Hedin,
  Rehr, and Bardyszewski}}]{campbell}
\bibinfo{author}{\bibfnamefont{L.}~\bibnamefont{Campbell}},
  \bibinfo{author}{\bibfnamefont{L.}~\bibnamefont{Hedin}},
  \bibinfo{author}{\bibfnamefont{J.~J.} \bibnamefont{Rehr}}, \bibnamefont{and}
  \bibinfo{author}{\bibfnamefont{W.}~\bibnamefont{Bardyszewski}},
  \bibinfo{journal}{Phys. Rev. B} \textbf{\bibinfo{volume}{65}},
  \bibinfo{pages}{064107} (\bibinfo{year}{2002}).

\bibitem[{\citenamefont{Fujikawa}(2001)}]{fujikawa}
\bibinfo{author}{\bibfnamefont{T.}~\bibnamefont{Fujikawa}},
  \bibinfo{journal}{J. Synchrotron Rad.} \textbf{\bibinfo{volume}{8}},
  \bibinfo{pages}{76} (\bibinfo{year}{2001}).

\bibitem[{\citenamefont{Lee et~al.}(1999)\citenamefont{Lee, Gunnarsson, and
  Hedin}}]{lgh}
\bibinfo{author}{\bibfnamefont{J.~D.} \bibnamefont{Lee}},
  \bibinfo{author}{\bibfnamefont{O.}~\bibnamefont{Gunnarsson}},
  \bibnamefont{and} \bibinfo{author}{\bibfnamefont{L.}~\bibnamefont{Hedin}},
  \bibinfo{journal}{Phys. Rev. B} \textbf{\bibinfo{volume}{60}},
  \bibinfo{pages}{8034} (\bibinfo{year}{1999}).

\bibitem[{\citenamefont{Hedin et~al.}(1998)\citenamefont{Hedin, Michiels, and
  Inglesfield}}]{hmi}
\bibinfo{author}{\bibfnamefont{L.}~\bibnamefont{Hedin}},
  \bibinfo{author}{\bibfnamefont{J.}~\bibnamefont{Michiels}}, \bibnamefont{and}
  \bibinfo{author}{\bibfnamefont{J.}~\bibnamefont{Inglesfield}},
  \bibinfo{journal}{Phys. Rev. B} \textbf{\bibinfo{volume}{58}},
  \bibinfo{pages}{15565} (\bibinfo{year}{1998}).

\bibitem[{\citenamefont{Klevak et~al.}(2014)\citenamefont{Klevak, Kas, and
  Rehr}}]{klevak14}
\bibinfo{author}{\bibfnamefont{E.}~\bibnamefont{Klevak}},
  \bibinfo{author}{\bibfnamefont{J.~J.} \bibnamefont{Kas}}, \bibnamefont{and}
  \bibinfo{author}{\bibfnamefont{J.~J.} \bibnamefont{Rehr}},
  \bibinfo{journal}{Phys. Rev. B} \textbf{\bibinfo{volume}{89}},
  \bibinfo{pages}{085123} (\bibinfo{year}{2014}).

\bibitem[{\citenamefont{Combescot and Nozieres}(1971)}]{NC}
\bibinfo{author}{\bibfnamefont{M.}~\bibnamefont{Combescot}} \bibnamefont{and}
  \bibinfo{author}{\bibfnamefont{P.}~\bibnamefont{Nozieres}},
  \bibinfo{journal}{J. de Physique} \textbf{\bibinfo{volume}{32}},
  \bibinfo{pages}{913} (\bibinfo{year}{1971}).

\bibitem[{\citenamefont{Tzavala et~al.}(2020)\citenamefont{Tzavala, Kas,
  Reining, and Rehr}}]{tzavala}
\bibinfo{author}{\bibfnamefont{M.}~\bibnamefont{Tzavala}},
  \bibinfo{author}{\bibfnamefont{J.~J.} \bibnamefont{Kas}},
  \bibinfo{author}{\bibfnamefont{L.}~\bibnamefont{Reining}}, \bibnamefont{and}
  \bibinfo{author}{\bibfnamefont{J.~J.} \bibnamefont{Rehr}},
  \bibinfo{journal}{Phys. Rev. Research} \textbf{\bibinfo{volume}{2}},
  \bibinfo{pages}{033147} (\bibinfo{year}{2020}).

\bibitem[{\citenamefont{Liang et~al.}(2017)\citenamefont{Liang, Vinson,
  Pemmaraju, Drisdell, Shirley, and Prendergast}}]{prendergast}
\bibinfo{author}{\bibfnamefont{Y.}~\bibnamefont{Liang}},
  \bibinfo{author}{\bibfnamefont{J.}~\bibnamefont{Vinson}},
  \bibinfo{author}{\bibfnamefont{S.}~\bibnamefont{Pemmaraju}},
  \bibinfo{author}{\bibfnamefont{W.~S.} \bibnamefont{Drisdell}},
  \bibinfo{author}{\bibfnamefont{E.~L.} \bibnamefont{Shirley}},
  \bibnamefont{and}
  \bibinfo{author}{\bibfnamefont{D.}~\bibnamefont{Prendergast}},
  \bibinfo{journal}{Phys. Rev. Lett.} \textbf{\bibinfo{volume}{118}},
  \bibinfo{pages}{096402} (\bibinfo{year}{2017}).

\bibitem[{\citenamefont{Biermann and van Roekeghem}(2016)}]{biermann2}
\bibinfo{author}{\bibfnamefont{S.}~\bibnamefont{Biermann}} \bibnamefont{and}
  \bibinfo{author}{\bibfnamefont{A.}~\bibnamefont{van Roekeghem}},
  \bibinfo{journal}{J. Electron Spectr. Relat. Phen.}
  \textbf{\bibinfo{volume}{208}}, \bibinfo{pages}{17} (\bibinfo{year}{2016}).

\bibitem[{\citenamefont{Bagus et~al.}(2022)\citenamefont{Bagus, Nelin, Brundle,
  Crist, Lahiri, and Rosso}}]{bagus2022}
\bibinfo{author}{\bibfnamefont{P.~S.} \bibnamefont{Bagus}},
  \bibinfo{author}{\bibfnamefont{C.~J.} \bibnamefont{Nelin}},
  \bibinfo{author}{\bibfnamefont{C.}~\bibnamefont{Brundle}},
  \bibinfo{author}{\bibfnamefont{B.~V.} \bibnamefont{Crist}},
  \bibinfo{author}{\bibfnamefont{N.}~\bibnamefont{Lahiri}}, \bibnamefont{and}
  \bibinfo{author}{\bibfnamefont{K.~M.} \bibnamefont{Rosso}},
  \bibinfo{journal}{Phys. Chem. Chem. Phys.} \textbf{\bibinfo{volume}{24}},
  \bibinfo{pages}{4562} (\bibinfo{year}{2022}).

\bibitem[{\citenamefont{Kas et~al.}(2015)\citenamefont{Kas, Vila, Rehr, and
  Chambers}}]{Kas2015}
\bibinfo{author}{\bibfnamefont{J.~J.} \bibnamefont{Kas}},
  \bibinfo{author}{\bibfnamefont{F.~D.} \bibnamefont{Vila}},
  \bibinfo{author}{\bibfnamefont{J.~J.} \bibnamefont{Rehr}}, \bibnamefont{and}
  \bibinfo{author}{\bibfnamefont{S.~A.} \bibnamefont{Chambers}},
  \bibinfo{journal}{Phys. Rev. B} \textbf{\bibinfo{volume}{91}},
  \bibinfo{pages}{121112} (\bibinfo{year}{2015}).

\bibitem[{\citenamefont{Woicik et~al.}(2020)\citenamefont{Woicik, Weiland,
  Jaye, Fischer, Rumaiz, Shirley, Kas, and Rehr}}]{woicik1}
\bibinfo{author}{\bibfnamefont{J.}~\bibnamefont{Woicik}},
  \bibinfo{author}{\bibfnamefont{C.}~\bibnamefont{Weiland}},
  \bibinfo{author}{\bibfnamefont{C.}~\bibnamefont{Jaye}},
  \bibinfo{author}{\bibfnamefont{D.}~\bibnamefont{Fischer}},
  \bibinfo{author}{\bibfnamefont{A.}~\bibnamefont{Rumaiz}},
  \bibinfo{author}{\bibfnamefont{E.}~\bibnamefont{Shirley}},
  \bibinfo{author}{\bibfnamefont{J.}~\bibnamefont{Kas}}, \bibnamefont{and}
  \bibinfo{author}{\bibfnamefont{J.}~\bibnamefont{Rehr}},
  \bibinfo{journal}{Phys. Rev. B} \textbf{\bibinfo{volume}{101}},
  \bibinfo{pages}{245119} (\bibinfo{year}{2020}).

\bibitem[{\citenamefont{Woicik et~al.}({2020})\citenamefont{Woicik, Weiland,
  Rumaiz, Brumbach, Ablett, Shirley, Kas, and Rehr}}]{woicik2}
\bibinfo{author}{\bibfnamefont{J.~C.} \bibnamefont{Woicik}},
  \bibinfo{author}{\bibfnamefont{C.}~\bibnamefont{Weiland}},
  \bibinfo{author}{\bibfnamefont{A.~K.} \bibnamefont{Rumaiz}},
  \bibinfo{author}{\bibfnamefont{M.~T.} \bibnamefont{Brumbach}},
  \bibinfo{author}{\bibfnamefont{J.~M.} \bibnamefont{Ablett}},
  \bibinfo{author}{\bibfnamefont{E.~L.} \bibnamefont{Shirley}},
  \bibinfo{author}{\bibfnamefont{J.~J.} \bibnamefont{Kas}}, \bibnamefont{and}
  \bibinfo{author}{\bibfnamefont{J.~J.} \bibnamefont{Rehr}},
  \bibinfo{journal}{{Phys. Rev. B}} \textbf{\bibinfo{volume}{{101}}},
  \bibinfo{pages}{245105} (\bibinfo{year}{{2020}}).

\bibitem[{\citenamefont{Kas et~al.}(2020)\citenamefont{Kas, Vila, and
  Rehr}}]{feff10}
\bibinfo{author}{\bibfnamefont{J.}~\bibnamefont{Kas}},
  \bibinfo{author}{\bibfnamefont{F.}~\bibnamefont{Vila}}, \bibnamefont{and}
  \bibinfo{author}{\bibfnamefont{J.}~\bibnamefont{Rehr}}
  (\bibinfo{year}{2020}).

\bibitem[{\citenamefont{Aryasetiawan et~al.}(1996)\citenamefont{Aryasetiawan,
  Hedin, and Karlsson}}]{aryasetiawan}
\bibinfo{author}{\bibfnamefont{F.}~\bibnamefont{Aryasetiawan}},
  \bibinfo{author}{\bibfnamefont{L.}~\bibnamefont{Hedin}}, \bibnamefont{and}
  \bibinfo{author}{\bibfnamefont{K.}~\bibnamefont{Karlsson}},
  \bibinfo{journal}{Phys. Rev. Lett.} \textbf{\bibinfo{volume}{77}},
  \bibinfo{pages}{2268} (\bibinfo{year}{1996}).

\bibitem[{\citenamefont{Guzzo et~al.}(2011)\citenamefont{Guzzo, Lani, Sottile,
  Romaniello, Gatti, Kas, Rehr, Silly, Sirotti, and Reining}}]{guzzo}
\bibinfo{author}{\bibfnamefont{M.}~\bibnamefont{Guzzo}},
  \bibinfo{author}{\bibfnamefont{G.}~\bibnamefont{Lani}},
  \bibinfo{author}{\bibfnamefont{F.}~\bibnamefont{Sottile}},
  \bibinfo{author}{\bibfnamefont{P.}~\bibnamefont{Romaniello}},
  \bibinfo{author}{\bibfnamefont{M.}~\bibnamefont{Gatti}},
  \bibinfo{author}{\bibfnamefont{J.~J.} \bibnamefont{Kas}},
  \bibinfo{author}{\bibfnamefont{J.~J.} \bibnamefont{Rehr}},
  \bibinfo{author}{\bibfnamefont{M.~G.} \bibnamefont{Silly}},
  \bibinfo{author}{\bibfnamefont{F.}~\bibnamefont{Sirotti}}, \bibnamefont{and}
  \bibinfo{author}{\bibfnamefont{L.}~\bibnamefont{Reining}},
  \bibinfo{journal}{Phys. Rev. Lett.} \textbf{\bibinfo{volume}{107}},
  \bibinfo{pages}{166401} (\bibinfo{year}{2011}).

\bibitem[{\citenamefont{Landau}(1944)}]{landau44}
\bibinfo{author}{\bibfnamefont{L.~D.} \bibnamefont{Landau}},
  \bibinfo{journal}{J. Phys.} \textbf{\bibinfo{volume}{8}},
  \bibinfo{pages}{201} (\bibinfo{year}{1944}).

\bibitem[{\citenamefont{Yabana and Bertsch}(1996)}]{Yabana-Bertsch}
\bibinfo{author}{\bibfnamefont{K.}~\bibnamefont{Yabana}} \bibnamefont{and}
  \bibinfo{author}{\bibfnamefont{G.~F.} \bibnamefont{Bertsch}},
  \bibinfo{journal}{Phys. Rev. B} \textbf{\bibinfo{volume}{54}},
  \bibinfo{pages}{4484} (\bibinfo{year}{1996}).

\bibitem[{\citenamefont{Kas et~al.}(2021)\citenamefont{Kas, Vila, Pemmaraju,
  Tan, and Rehr}}]{kas2021feff10}
\bibinfo{author}{\bibfnamefont{J.}~\bibnamefont{Kas}},
  \bibinfo{author}{\bibfnamefont{F.}~\bibnamefont{Vila}},
  \bibinfo{author}{\bibfnamefont{C.}~\bibnamefont{Pemmaraju}},
  \bibinfo{author}{\bibfnamefont{T.}~\bibnamefont{Tan}}, \bibnamefont{and}
  \bibinfo{author}{\bibfnamefont{J.}~\bibnamefont{Rehr}}, \bibinfo{journal}{J.
  Synchrotron Rad.} \textbf{\bibinfo{volume}{28}} (\bibinfo{year}{2021}).

\bibitem[{\citenamefont{Almbladh and Hedin}(1983)}]{almbladh}
\bibinfo{author}{\bibfnamefont{C.-O.} \bibnamefont{Almbladh}} \bibnamefont{and}
  \bibinfo{author}{\bibfnamefont{L.}~\bibnamefont{Hedin}},
  \bibinfo{journal}{1b, Hrsg. E. Koch (North Holland, Amsterdam 1983) S} pp.
  \bibinfo{pages}{607--904} (\bibinfo{year}{1983}).

\bibitem[{\citenamefont{Ankudinov et~al.}(2005)\citenamefont{Ankudinov,
  Takimoto, and Rehr}}]{tddft-bse}
\bibinfo{author}{\bibfnamefont{A.~L.} \bibnamefont{Ankudinov}},
  \bibinfo{author}{\bibfnamefont{Y.}~\bibnamefont{Takimoto}}, \bibnamefont{and}
  \bibinfo{author}{\bibfnamefont{J.~J.} \bibnamefont{Rehr}},
  \bibinfo{journal}{Phys. Rev. B} \textbf{\bibinfo{volume}{71}},
  \bibinfo{pages}{165110} (\bibinfo{year}{2005}).

\bibitem[{\citenamefont{Prange et~al.}(2009)\citenamefont{Prange, Rehr, Rivas,
  Kas, and Lawson}}]{prange_opcons}
\bibinfo{author}{\bibfnamefont{M.~P.} \bibnamefont{Prange}},
  \bibinfo{author}{\bibfnamefont{J.~J.} \bibnamefont{Rehr}},
  \bibinfo{author}{\bibfnamefont{G.}~\bibnamefont{Rivas}},
  \bibinfo{author}{\bibfnamefont{J.~J.} \bibnamefont{Kas}}, \bibnamefont{and}
  \bibinfo{author}{\bibfnamefont{J.~W.} \bibnamefont{Lawson}},
  \bibinfo{journal}{Phys. Rev. B} \textbf{\bibinfo{volume}{80}},
  \bibinfo{pages}{155110} (\bibinfo{year}{2009}).

\bibitem[{\citenamefont{Zangwill and Soven.}(1980)}]{zangsov}
\bibinfo{author}{\bibfnamefont{A.}~\bibnamefont{Zangwill}} \bibnamefont{and}
  \bibinfo{author}{\bibfnamefont{P.}~\bibnamefont{Soven.}},
  \bibinfo{journal}{Phys. Rev. A} \textbf{\bibinfo{volume}{21}},
  \bibinfo{pages}{1561} (\bibinfo{year}{1980}).

\bibitem[{\citenamefont{Stott and Zaremba}(1980)}]{zaremba}
\bibinfo{author}{\bibfnamefont{M.~J.} \bibnamefont{Stott}} \bibnamefont{and}
  \bibinfo{author}{\bibfnamefont{E.}~\bibnamefont{Zaremba}},
  \bibinfo{journal}{Phys. Rev. A} \textbf{\bibinfo{volume}{21}},
  \bibinfo{pages}{12} (\bibinfo{year}{1980}).

\bibitem[{\citenamefont{Shirley and Martin}(1993)}]{shirleyatom}
\bibinfo{author}{\bibfnamefont{E.~L.} \bibnamefont{Shirley}} \bibnamefont{and}
  \bibinfo{author}{\bibfnamefont{R.~M.} \bibnamefont{Martin}},
  \bibinfo{journal}{Phys. Rev. B} \textbf{\bibinfo{volume}{47}},
  \bibinfo{pages}{15404} (\bibinfo{year}{1993}).

\bibitem[{\citenamefont{Jr.}(1976)}]{norman}
\bibinfo{author}{\bibfnamefont{J.~G.~N.} \bibnamefont{Jr.}},
  \bibinfo{journal}{Molecular Physics} \textbf{\bibinfo{volume}{31}},
  \bibinfo{pages}{1191} (\bibinfo{year}{1976}).

\bibitem[{\citenamefont{Desclaux}(1975)}]{Descl}
\bibinfo{author}{\bibfnamefont{J.~P.} \bibnamefont{Desclaux}},
  \bibinfo{journal}{Comp. Phys. Comm.} \textbf{\bibinfo{volume}{9}},
  \bibinfo{pages}{31} (\bibinfo{year}{1975}).

\bibitem[{\citenamefont{Ankudinov}(1996)}]{alex}
\bibinfo{author}{\bibfnamefont{A.~L.} \bibnamefont{Ankudinov}}, Ph.D. thesis,
  \bibinfo{school}{University of {W}ashington} (\bibinfo{year}{1996}).

\bibitem[{\citenamefont{Noz\`ieres and De~Dominicis}(1969)}]{ND}
\bibinfo{author}{\bibfnamefont{P.}~\bibnamefont{Noz\`ieres}} \bibnamefont{and}
  \bibinfo{author}{\bibfnamefont{C.~T.} \bibnamefont{De~Dominicis}},
  \bibinfo{journal}{Phys. Rev.} \textbf{\bibinfo{volume}{178}},
  \bibinfo{pages}{1097} (\bibinfo{year}{1969}).

\bibitem[{\citenamefont{Rehr et~al.}(2020)\citenamefont{Rehr, Vila, Kas,
  Hirshberg, Kowalski, and Peng}}]{rehr20}
\bibinfo{author}{\bibfnamefont{J.~J.} \bibnamefont{Rehr}},
  \bibinfo{author}{\bibfnamefont{F.~D.} \bibnamefont{Vila}},
  \bibinfo{author}{\bibfnamefont{J.~J.} \bibnamefont{Kas}},
  \bibinfo{author}{\bibfnamefont{N.~Y.} \bibnamefont{Hirshberg}},
  \bibinfo{author}{\bibfnamefont{K.}~\bibnamefont{Kowalski}}, \bibnamefont{and}
  \bibinfo{author}{\bibfnamefont{B.}~\bibnamefont{Peng}}, \bibinfo{journal}{J.
  Chem. Phys.} \textbf{\bibinfo{volume}{152}}, \bibinfo{pages}{174113}
  (\bibinfo{year}{2020}).

\bibitem[{\citenamefont{Shirley}(1972)}]{shirleyBG}
\bibinfo{author}{\bibfnamefont{D.~A.} \bibnamefont{Shirley}},
  \bibinfo{journal}{Phys. Rev. B} \textbf{\bibinfo{volume}{5}},
  \bibinfo{pages}{4709} (\bibinfo{year}{1972}).

\bibitem[{\citenamefont{Lundqvist}(1967)}]{LundqvistII}
\bibinfo{author}{\bibfnamefont{B.~I.} \bibnamefont{Lundqvist}},
  \bibinfo{journal}{Phys. kondens. Materie} \textbf{\bibinfo{volume}{6}},
  \bibinfo{pages}{206} (\bibinfo{year}{1967}).

\bibitem[{sup()}]{suppmat}
\bibinfo{note}{See Supplemental Material at [URL will be inserted by publisher]
  for derivation of the screened potentias within the plasmon pole model.}

\bibitem[{\citenamefont{Citrin et~al.}(1977)\citenamefont{Citrin, Wertheim, and
  Baer}}]{citrin1977}
\bibinfo{author}{\bibfnamefont{P.~H.} \bibnamefont{Citrin}},
  \bibinfo{author}{\bibfnamefont{G.~K.} \bibnamefont{Wertheim}},
  \bibnamefont{and} \bibinfo{author}{\bibfnamefont{Y.}~\bibnamefont{Baer}},
  \bibinfo{journal}{Phys. Rev. B} \textbf{\bibinfo{volume}{16}},
  \bibinfo{pages}{4256} (\bibinfo{year}{1977}).

\bibitem[{\citenamefont{Bocquet et~al.}(1996)\citenamefont{Bocquet, Mizokawa,
  Morikawa, Fujimori, Barman, Maiti, Sarma, Tokura, and Onoda}}]{bocquet1996}
\bibinfo{author}{\bibfnamefont{A.~E.} \bibnamefont{Bocquet}},
  \bibinfo{author}{\bibfnamefont{T.}~\bibnamefont{Mizokawa}},
  \bibinfo{author}{\bibfnamefont{K.}~\bibnamefont{Morikawa}},
  \bibinfo{author}{\bibfnamefont{A.}~\bibnamefont{Fujimori}},
  \bibinfo{author}{\bibfnamefont{S.~R.} \bibnamefont{Barman}},
  \bibinfo{author}{\bibfnamefont{K.}~\bibnamefont{Maiti}},
  \bibinfo{author}{\bibfnamefont{D.~D.} \bibnamefont{Sarma}},
  \bibinfo{author}{\bibfnamefont{Y.}~\bibnamefont{Tokura}}, \bibnamefont{and}
  \bibinfo{author}{\bibfnamefont{M.}~\bibnamefont{Onoda}},
  \bibinfo{journal}{Phys. Rev. B} \textbf{\bibinfo{volume}{53}},
  \bibinfo{pages}{1161} (\bibinfo{year}{1996}).

\bibitem[{\citenamefont{Vast et~al.}(2002)\citenamefont{Vast, Reining, Olevano,
  Schattschneider, and Jouffrey}}]{vast2002}
\bibinfo{author}{\bibfnamefont{N.}~\bibnamefont{Vast}},
  \bibinfo{author}{\bibfnamefont{L.}~\bibnamefont{Reining}},
  \bibinfo{author}{\bibfnamefont{V.}~\bibnamefont{Olevano}},
  \bibinfo{author}{\bibfnamefont{P.}~\bibnamefont{Schattschneider}},
  \bibnamefont{and} \bibinfo{author}{\bibfnamefont{B.}~\bibnamefont{Jouffrey}},
  \bibinfo{journal}{Phys. Rev. Lett.} \textbf{\bibinfo{volume}{88}},
  \bibinfo{pages}{037601} (\bibinfo{year}{2002}).

\bibitem[{\citenamefont{Miyakawa}(1968)}]{miyakawa1968}
\bibinfo{author}{\bibfnamefont{T.}~\bibnamefont{Miyakawa}},
  \bibinfo{journal}{J. Phys. Soc. Jpn.} \textbf{\bibinfo{volume}{24}},
  \bibinfo{pages}{768} (\bibinfo{year}{1968}).

\bibitem[{\citenamefont{\v{Z}upanovi\'{c}
  et~al.}(1997)\citenamefont{\v{Z}upanovi\'{c}, Bjeli\v{s}, and
  Bari\v{s}i\'{c}}}]{zupanovic1997}
\bibinfo{author}{\bibfnamefont{P.}~\bibnamefont{\v{Z}upanovi\'{c}}},
  \bibinfo{author}{\bibfnamefont{A.}~\bibnamefont{Bjeli\v{s}}},
  \bibnamefont{and}
  \bibinfo{author}{\bibfnamefont{S.}~\bibnamefont{Bari\v{s}i\'{c}}},
  \bibinfo{journal}{Z. Phys. B} \textbf{\bibinfo{volume}{101}},
  \bibinfo{pages}{397–404} (\bibinfo{year}{1997}).

\bibitem[{\citenamefont{Gunn and Inkson}(1979)}]{gunn1979}
\bibinfo{author}{\bibfnamefont{J.~M.~F.} \bibnamefont{Gunn}} \bibnamefont{and}
  \bibinfo{author}{\bibfnamefont{J.~C.} \bibnamefont{Inkson}},
  \bibinfo{journal}{J. Phys. C: Solid State Phys.}
  \textbf{\bibinfo{volume}{12}}, \bibinfo{pages}{1049} (\bibinfo{year}{1979}).

\bibitem[{\citenamefont{de~Boer et~al.}(1984)\citenamefont{de~Boer, Haas, and
  Sawatzky}}]{deboer}
\bibinfo{author}{\bibfnamefont{D.~K.~G.} \bibnamefont{de~Boer}},
  \bibinfo{author}{\bibfnamefont{C.}~\bibnamefont{Haas}}, \bibnamefont{and}
  \bibinfo{author}{\bibfnamefont{G.~A.} \bibnamefont{Sawatzky}},
  \bibinfo{journal}{Phys. Rev. B} \textbf{\bibinfo{volume}{29}},
  \bibinfo{pages}{4401} (\bibinfo{year}{1984}).

\bibitem[{\citenamefont{Cudazzo and Reining}(2020)}]{cudazzo}
\bibinfo{author}{\bibfnamefont{P.}~\bibnamefont{Cudazzo}} \bibnamefont{and}
  \bibinfo{author}{\bibfnamefont{L.}~\bibnamefont{Reining}},
  \bibinfo{journal}{Phys. Rev. Research} \textbf{\bibinfo{volume}{2}},
  \bibinfo{pages}{012032} (\bibinfo{year}{2020}).

\end{thebibliography}

\section{Supplementary Material: Plasmon-pole model}
\begin{equation}
    \epsilon^{-1}(q,\omega) = 1 + \frac{\omega_p^2}{(\omega+i\delta)^2-\omega_q^2}.
\end{equation}

\begin{equation}
    W(q,\omega) = \frac{4\pi}{q^2}\epsilon^{-1}(q,\omega).
\end{equation}

\begin{align}
    W(r,\omega) &= \int\ \frac{d^3q}{(2\pi)^3}\frac{4\pi}{q^2}\epsilon^{-1}(q,\omega)e^{i\bf{q}\cdot\bf{r}} \nonumber \\
    & = \frac{2}{\pi}\int_0^\infty dq\ \epsilon^{-1}(q,\omega)\frac{\sin(q r)}{q r} 
\end{align}

Expanding $\sin(qr)=[\exp(iqr)-\exp(-iqr)]/2i$, and noting that $\epsilon^{-1}$ is an even function of $q$ gives,
\begin{equation}
    W(r,\omega) = \frac{1}{i \pi} \int_{-\infty}^{\infty}\epsilon^{-1}(q,\omega)\frac{e^{iqr}}{qr}
\end{equation}

The first term in $W$ coming from the $1$ in $\epsilon^{-1}$ gives the unscreened Coulomb potential $v(r) = 1/r$. The second term $\tilde W = W-v$ is given by,
\begin{equation}
    \tilde W(r,\omega) = \frac{1}{i\pi}\int_{-\infty}^{\infty}dq\ \frac{e^{iqr}}{qr}\frac{\omega_p^2}{(\omega-\omega_q+i\delta)(\omega+\omega_q+i\delta)}.
\end{equation}
For simplicity, we take the dispersion relation $\omega_q=\omega_p + q^2/2$, which gives an integrand with five poles at $q=0$, 
$q=\pm \sqrt{2(\omega-\omega_p+i\delta)}$, 
and $q=\pm i\sqrt{2(\omega+\omega_p)}$. The appearance of $\exp(iqr)$ in the requires that the contour be closed in the upper half plane, so that only the poles with positive imaginary part, as well as the pole on the real axis contribute. Assuming positive frequency only, we have poles at $q=0$, $q_1=\sqrt{2(\omega-\omega_p+i\delta)}$, and $q_2=i\sqrt{2(\omega+\omega_p)}$. The $q=0$ pole gives a frequency dependent contribution $\tilde W_0$ proportional to the unscreened Coulomb interaction,
\begin{equation}
    \tilde W_0(r,\omega) = \frac{1}{r}\frac{\omega_p^2}{(\omega+i\delta)^2-\omega_p^2}.
\end{equation}

In the zero frequency limit, this term cancels the $v(r)$ in $W$, which is necessary if the screened Coulomb interaction is going to fall off faster than $1/r$ at large $r$.
The pole at $q_1$ gives a contribution $\tilde W_1$ which is real and exponentially decaying for $\omega < \omega_p$, and complex and oscillatory for $\omega > \omega_p$, i.e., 
\begin{equation}
    \tilde W_1(q,\omega) = -\frac{\omega_p^2}{\omega q_1}\frac{e^{i q_1 r}}{q_1 r}.
\end{equation}
The above term is also the only term that is complex, 
so that one can immediately find the quasiboson excitation spectrum by taking the limit of the imaginary part as $r \to 0$, 
\begin{align}
    \beta(\omega) &= -\frac{1}{\pi}\lim_{r \to 0} \frac{\omega_p^2}{\omega q_1}\frac{\sin(q_1 r)}{q_1 r}\theta(\omega - \omega_p) \nonumber \\
    &= \frac{1}{\pi}\frac{\omega_p^2}{\omega q_1}\theta(\omega - \omega_p),
\end{align}
which matches the result found previously for the same model.\cite{hedin99rev} Note that the $q=q_1$ contribution becomes unphysically large at low frequency, and is singular at zero frequency, and this is precisely where the $q=q_2$ term is important. This gives a contribution $\tilde W_2$,
\begin{equation}
    \tilde W_2(r,\omega) = \frac{\omega_p^2}{\omega q_2}\frac{e^{iq_2r}}{q_2r}
\end{equation}
At low frequency $\omega < \omega_p$, $q_1$ becomes imaginary, so that $\tilde W_1$ and $\tilde W_2$ tend to cancel,
\begin{align}
    \tilde W_1(r,\omega) &+ \tilde W_2(r,\omega) \nonumber \\
    &= \frac{\omega_p^2}{\omega r}\left[ \frac{e^{-\kappa_1 r}}{\kappa_1^2} - \frac{e^{-\kappa_2 r}}{\kappa_2^2}\right] \nonumber,
\end{align}
where $\kappa_1 = \sqrt{2(\omega_p-\omega)}$, and $\kappa_2 = \sqrt{2(\omega_p+\omega)}$, $\kappa_0=\sqrt{2\omega_p}$, $\delta\kappa=\omega/\kappa_0$.
Taking the zero frequency limit of the above gives the zero frequency limit of $W$, since $v-W_0=0$ at zero frequency, 
\begin{equation}
    \lim_{\omega \to 0} W(r,\omega) = = e^{-\kappa_0 r}\left[\frac{\kappa_0}{2} + \frac{1}{r} \right].
\end{equation}
Thus in its entirety, the screened potential is,
\begin{align}
    W(r,\omega) &= \frac{1}{r}\left[1+\frac{\omega_p^2}{(\omega+i\delta)^2-\omega_p^2}\right] \\
    &-\frac{\omega_p^2}{\omega}\frac{e^{-\kappa_2 r}}{\kappa_2^2 r} - \frac{\omega_p^2}{\omega}\frac{e^{i q_1 r}}{q_1^2 r}.
\end{align}
Finally, we can take real and imaginary parts of this function to diagnose its behavior at the plasmon frequency. 
\begin{align}
    {\rm Re}[W(r,\omega)] &= \frac{1}{r}\left\{ 1+\frac{\omega_p^2}{(\omega + \omega_p)(\omega-\omega_p)}\right.\nonumber \\
    & - \frac{\omega_p^2}{2\omega}\left[\frac{e^{-\kappa_2 r}}{\omega+\omega_p}-\frac{e^{-\kappa_1 r}}{\omega_p-\omega}\theta(\omega_p-\omega)\right. \nonumber \\
    &\left.\left.+\frac{\cos(q_1 r)}{\omega-\omega_p}\theta(\omega-\omega_p)\right] \right\}, \\
    {\rm Im}[W(r,\omega)] &= \frac{\omega_p^2}{r}\left[-\frac{1}{\omega+\omega_p}\delta(\omega-\omega_p)\right. \nonumber \\
    &\left.+\frac{\cos(q_1 r)}{2\omega}\delta(\omega-\omega_p) -\frac{\sin(q_1 r)}{\omega q_1^2}\theta(\omega-\omega_p)\right]\nonumber \\
    &=-\frac{\omega_p^2}{\omega q_1}\frac{\sin(q_1 r)}{q_1 r}\theta(\omega-\omega_p).
\end{align}
 Reiterating, $W(r,\omega)$ is real for frequencies less than the plasmon frequency $\omega < \omega_p$. At very low frequencies, the potential falls off exponentially, although a Yukawa term is also present, with a growing Coulombic term as frequency increases. At frequencies above the plasmon frequency, the real part also includes an oscillatory term $\sim\cos(q_1 r)/q_1 r$, and an imaginary part appears, which oscillates as $\sim\sin(q_1 r)/q_1 r$. Near $\omega = \omega_p$ is also interesting to analyze. At the plasmon frequency, the second term in the real part cancels with the last two terms, giving ${\rm Re}W \sim 1/r$ at large r, while the imaginary part is singular, with a behavior ${\rm Im}W\sim 1/\sqrt{\omega-\omega_p}\theta(\omega-\omega_p)$. This means that somewhere near the plasmon frequency, the oscillations in $W$ become very large, and at high frequency die down, similar to what is seen in Fig. 2. Finally, in the high frequency limit, all terms except the bare Coulomb potential scale as $\sim 1/\omega^2$, so that we essentially retain the bare Coulomb potential, as expected. 
\end{document}